\documentclass{article}

\usepackage{arxiv}

\usepackage[utf8]{inputenc} 
\usepackage[T1]{fontenc}    
\usepackage{hyperref}       
\usepackage{url}            
\usepackage{booktabs}       
\usepackage{amsfonts}       
\usepackage{nicefrac}       
\usepackage{microtype}      
\usepackage{graphicx}
\usepackage{amssymb}
\usepackage{amsmath}
\usepackage{lineno}
\usepackage{hyperref}
\usepackage{listings}
\usepackage{subfigure}
\usepackage{comment}
\usepackage{enumitem}
\usepackage[dvipsnames]{xcolor}

\title{A Joint Inversion-Segmentation approach to Assisted Seismic Interpretation}

\author{
  Matteo Ravasi \\
  KAUST\\
  Thuwal, Kingdom of Saudi Arabia \\
  \texttt{matteo.ravasi@kaust.edu.sa} \\
   \And
 Claire Birnie \\
  KAUST\\
  Thuwal, Kingdom of Saudi Arabia \\
  \texttt{cebirnie@gmail.com}}
  
\begin{document}

\chead{Joint Inversion-Segmentation for Seismic Interpretation}

\maketitle

\begin{abstract}
  Structural seismic interpretation and quantitative characterization are historically intertwined processes. The latter provides estimates of properties of the subsurface which can be used to aid structural interpretation alongside the original seismic data and a number of other seismic attributes. In this work, we redefine this process as a inverse problem which tries to jointly estimate subsurface properties (i.e., acoustic impedance) and a piece-wise segmented representation of the subsurface based on user-defined \textit{macro-classes}. By inverting for the quantities simultaneously, the inversion is primed with prior knowledge about the regions of interest, whilst at the same time it constrains this belief with the actual seismic measurements. As the proposed functional is separable in the two quantities, these are optimized in an alternating fashion, where each subproblem is solved using a Primal-Dual algorithm. Subsequently, each class is used as input to a workflow which aims to extract the perimeter of the detected shapes and to produce unique horizons. The effectiveness of the proposed method is illustrated through numerical examples on synthetic and field datasets.
\end{abstract}

\section{Introduction}
Seismic interpretation forms part of an integrated workflow for reservoir mapping and characterization, and represents one of the most important steps for the successful exploration of underground resources and management of existing ones. It is traditionally divided into two components: structural and quantitative interpretation. The former aims at identifying the geological framework (i.e., horizons and faults) whilst the latter focuses on retrieving properties of the subsurface. 

Whilst structural interpretation is nowadays mostly carried out manually and led by geological understanding, several key geophysical principles are also followed early on to identify an interpretation strategy to guide to the interpreter's job: with the aid of sonic and density logs, a first step in an interpretation project is represented by the creation of well-ties, which allow interpreters to link the geological stratigraphy to the observed seismic response - in other words, interpreters look for changes in acoustic and shear impedances (or velocity ratios) in log data, which trigger a seismic response, and perform seismic modelling to identify such events in the observed data. However, as the word implies, \textit{interpretation} is a subjective matter whose ultimate goal is that of reconstructing the geological story contained in a seismic volume \cite{Herron}.

With recent advances in artificial intelligence and deep learning, several attempts at automatic seismic interpretation have been reported \cite{Waldeland, Haibin, Shi1, Shi2}. A common feature of most of these approaches is that of turning a geologically-driven process into a computer vision and pattern recognition problem. They have been shown to be very successful for some specific tasks where a clear pattern can be observed and learned - e.g., for top salt interpretation \cite{Waldeland}, where a seismic rich area is abruptly interrupted by a strong reflector, followed by an area with little to no seismic signal). On the other hand, these methods currently fail at data integration, something that human interpreters are very good at. The use of well-ties is a good example of this - where well-tie principles are not directly encoded in the machine learning training process resulting in the need for interpreters inputs prior to training. Typically, interpreters are still required to define the horizons of interest and in some cases interpret it for a number of lines in order to train a machine that is subsequently able to \textit{auto-track} the same horizons of interest in the rest of the dataset \cite{Haibin}. 

Post- and pre-stack inversion represent the workhorses in the process of quantitative characterization of the subsurface. Whilst coming in many different flavors, most commonly used methods estimate parameters by optimizing a least-squares functional with spatial regularization terms and, in the pre-stack case, extra regularization terms aimed at enforcing correlation between parameters \cite{Hampson}. However, since the input data is band-limited in nature and lacks both the very low and very high frequencies, least-squares approaches suffer from the inability of the inversion to fully recover the entire spectrum, leading to unwanted oscillatory behaviors near interfaces in the recovered model. While a properly defined low frequency background model can partially fill such a frequency gap, the remaining missing frequencies can only be compensated by providing additional prior information in the form of more suitable regularizers. TV regularization is one such type of regularization as it favours blockiness in the recovered impedance model; adding such regularization does however come at the cost of making the overall functional non-smooth and when used alone is known to lead to a systematic loss of contrast \cite{Osher}. The Split-Bregman algorithm \cite{Goldstein} has been successfully employed in solving L1-regularized problems in a variety of fields and it is used by \cite{Gholami} in the context of seismic impedance inversion of reflectivity data (after a preliminary step of blind deconvolution). Similarly, \cite{Wang} and \cite{Kolbjornsen} use the Alternating Direction Method of Multipliers (ADMM) for both post- and pre-stack inversion of band-limited data.

In this work, we propose a different approach to computer-aided seismic interpretation, which aims to reconcile the classical principles of seismic interpretation to the underlying optimization problem that we define and solve. The interpreter input is here limited to the definition of acoustic zones identified by a single acoustic impedance (AI) value (or elastic zones identified by triplets of parameters) that we wish to ultimately extract from the seismic data. A joint inversion and segmentation of the input seismic data is then performed and both inverted acoustic (or elastic) properties and seismic horizons are produced. Examples on synthetic and field datasets show that our methodology can successfully interpret key structures as defined by the interpreter in a fully automatic manner without any manual labelling or expensive upfront training. Moreover, as a by-product, both an acoustic impedance model and a segmentation model are retrieved that can be used to further refine the interpreted horizons as well as inputs to subsequent steps of reservoir modelling.

\section{A philosophical view of the problem}
Traditionally, seismic data has been interpreted by looking at a variety of attributes (e.g., full stack, partial stacks, seismic attributes) and most commercial softwares are built around this principle. Whilst used in some cases to aid interpretation, elastic properties derived from seismic pre- or post-stack inversion are generally not considered the main source of information against which to interpret horizons. Nevertheless, these properties are the physical reason behind what is observed within the seismic data: a large increase (or decrease) in acoustic impedance at a geological boundary produces a strong seismic response; conversely, a small change cannot trigger a response and so our ability to interpret such geological boundary is in vain. We therefore ask ourselves whether a machine, unhindered by human habits or limitations of current softwares, could be more successful at aiding interpretation if working directly with elastic properties other than the original seismic data? Or in other words, can we turn the problem of interpreting an horizon from following wiggles on seismic sections into a segmentation task for the acoustic (or elastic) properties in the geological formations of interest? In the next section we propose a mathematical formulation that translates into equations our objective.

\section*{Mathematical formulation}
Inspired by the work of \cite{Corona} on joint reconstruction and segmentation in the context of medical imaging, we define a functional to jointly invert seismic data for their acoustic impedance as well as to produce a segmentation of the subsurface into $N_c$ classes, each of which is characterized by a different range of impedance values. The underlying idea of inverting for these two parameters simultaneously is to inform the inversion with prior knowledge about the regions of interest, whilst at the same time constraining this belief with the seismic measurements. The functional for our specific problem can be written as:
\begin{equation}
\label{eq:joint}
(\mathbf{m}, \mathbf{V}) = \underset{\mathbf{m}, \mathbf{V} \in C} {\mathrm{argmin}}  \quad \frac{1}{2} ||\mathbf{d} - \mathbf{Gm}||_2^2 + \alpha TV(\mathbf{m})  + \delta \sum_{j=1}^{N_c} \sum_{i=1}^{N_x N_z}  V_{ji} (m_i - c_j)^2 +\beta \sum_{j=1}^{N_c}TV(\mathbf{V}^T_j)
\end{equation}
where $\mathbf{m}$ is a vector of size $N_x N_z \times 1$ that contains the natural logarithm of the acoustic impedance values in the area of the subsurface of interest, $\mathbf{V}$ is a matrix of size $ N_c \times N_xN_z$ whose columns contain the probability of each point in the subsurface to belong to each class, and $\mathbf{c}$ is a vector of size $N_c \times 1$ which contains the acoustic impedance values associated to each class. $\mathbf{G}=\mathbf{W}\mathbf{D}$ is the post-stack modelling operator composed of a first derivative operator $\mathbf{D}$ and a convolution operator $\mathbf{W}$ whose convolution kernel is the estimated wavelet $w$ divided by 2, and $\mathbf{d}$ is post-stack seismic data of size size $N_x N_z \times 1$. Here, we use the convention that $x_i$ is the i-th element of a vector and $\mathbf{X}_i$ represents the extraction of the i-th column of a matrix (whilst $\mathbf{X}^T_j$ is the j-th row of a matrix transposed into a column vector).

Furthermore, we further constrain each column of $\mathbf{V}$ to be in the unit Simplex, $C=\{ \mathbf{V}_i \in \mathbb{R}^{+}:  \sum_{j=1}^{N_c} V_{ji}=1 \} \quad \forall i=1,2,..., N_zN_x$. Note that by ensuring that the sum of the elements of each column of $\mathbf{V}$ is equal to 1 and every element is positive, we can interpret such values as the probability of each point in the subsurface belonging to a certain class. The Total Variation (TV) regularization term is defined as $TV(\mathbf{x})=||\nabla \mathbf{x}||_{2,1} = \sum_{i=1}^{N_xN_z} \sqrt{(\mathbf{x}_x)_i^2 +(\mathbf{x}_z)_i^2}$ where $\nabla:\mathbb{R}^{N_xN_z} \rightarrow \mathbb{R}^{2 \times  N_xN_z}$ is the gradient operator that transforms a vector into a matrix whose rows contain the derivatives $(\mathbf{x}_x)$ and $(\mathbf{x}_z)$ computed along the x- and z directions, respectively. Finally, $\alpha$, $\beta$, and $\delta$ are regularization parameters used to balance the inversion and segmentation terms of the functional.

\subsection{Alternating minimization}
In order to solve the functional in equation \ref{eq:joint} we note that, whilst non-convex in the joint argument $(\mathbf{m}, \mathbf{V})$, this cost function is convex in each individual variable space. A splitting approach is therefore used to minimise the two convex problems in an alternating fashion. Moreover, since solving TV-regularized problems can lead to a systematic loss of contrast in the estimated model \cite{Benning}, we replace both TV regularization terms with their generalized Bregman distance ($D_{TV}^{\mathbf{p}}(\mathbf{u}, \mathbf{u}') = TV(\mathbf{u}) - TV(\mathbf{u}') - (\mathbf{u}-\mathbf{u}')^T \mathbf{p}$) and Bregman iterations are introduced to solve each independent minimisation problem \cite{Osher}. The overall algorithm reads as:
\begin{subequations}
\label{eq:alg}
\begin{align}
& \mathbf{m}^k = \underset{\mathbf{m}} {\mathrm{argmin}} \quad \frac{1}{2} ||\mathbf{d} - \mathbf{Gm}||_2^2 + \alpha (TV(\mathbf{m}) - \mathbf{m}^T \mathbf{p}^{k-1}) + \delta \sum_{j=1}^{N_c} \sum_{i=1}^{N_x N_z} V^{k-1}_{ji} (m_i - c_j)^2 \\
& \mathbf{p}^k = \mathbf{p}^{k-1} - \frac{1}{\alpha} \left( \mathbf{G}^T(\mathbf{G}\mathbf{m}^k - \mathbf{d}) + 2\delta \sum_{j=1}^{N_c} \mathbf{V}^T_{j}(\mathbf{m}^k - c_j) \right) \\
& \mathbf{V}^k = \underset{\mathbf{V} \in C} {\mathrm{argmin}} \quad \delta \sum_{j=1}^{N_c} \sum_{i=1}^{N_x N_z} V_{ji} (m^k_i - c_j)^2 + \beta \sum_{j=1}^{N_c} (TV(\mathbf{V}_j^T) - (\mathbf{V}_j^T)^T \mathbf{Q}_j^{Tk-1})\\
& Q_{ji}^k = Q_{ji}^{k-1} - \frac{\delta}{\beta} (m^k_i - c_j)^2 \quad \forall j=1,2,..,N_c \quad i=1,2,..,N_xN_z
\end{align}
\end{subequations}
where $\mathbf{p}$ and $\mathbf{Q}$ are the sub-gradients of their corresponding Bregman distances. In the following, we will focus on the solution of equations 2a and 2c.

\subsection*{The Primal-Dual solver}
Proximal algorithms are a family of solvers able to minimize non-smooth, convex functionals like those in equations 2a and 2c \cite{Parikh}. Minimisation is performed in an iterative fashion and each iteration relies on the ability of evaluating the so-called \textit{proximal operator} of the functional $f$ to be minimised:
\begin{equation}
\label{eq:proximal}
prox_{\tau f}(\mathbf{u}) = \underset{\mathbf{x}} {\mathrm{argmin}}  \quad f(\mathbf{x}) + \frac{1}{2 \tau} ||\mathbf{x} - \mathbf{u}||_2^2
\end{equation}
As we will see in the following, closed form solutions exist for many commonly used functionals, meaning that these sub-problems can be solved very quickly with fast specialized methods.
 
The Primal-Dual algorithm \cite{Chambolle} is a special type of proximal solver that shows $O(1/N)$ rate of convergence in finite dimensions. It works by recasting any functional of this kind:
\begin{equation}
\label{eq:primal}
\underset{\mathbf{x}} {\mathrm{argmin}}  \quad f(\mathbf{K}\mathbf{x}) + \sum_i \mathbf{x}^T \mathbf{z}_i + g(\mathbf{x})
\end{equation}
into its primal-dual equivalent (a saddle-point problem):
\begin{equation}
\label{eq:primaldual}
\underset{\mathbf{x}} {\mathrm{argmin}} \; \underset{\mathbf{y}} {\mathrm{argmax}}  \quad \mathbf{y}^T (\mathbf{K}\mathbf{x}) + \sum_i \mathbf{z}_i^T \mathbf{x} + g(\mathbf{x}) - f^*(\mathbf{y})
\end{equation}
where $f$ and $g$ are convex (possibly non-smooth) functionals, $f^*$ is the convex conjugate of $f$ and $\mathbf{K}$ is a linear operator that maps the vector $\mathbf{x} \in R^m$ into a vector $\mathbf{y} \in R^n$.

A series of iterations is then introduced in order to obtain convergence at the saddle point:
\begin{equation}
\label{eq:pdalg}
\begin{cases}
    \mathbf{y}^{k+1} = prox_{\mu g^*}(\mathbf{y}^{k} +
    \mu \mathbf{K}\bar{\mathbf{x}}^{k})\\
    \mathbf{x}^{k+1} = prox_{\tau f}(\mathbf{x}^{k} -
    \tau (\mathbf{K}^H \mathbf{y}^{k+1} + \sum_i \mathbf{z}_i)) \\
    \bar{\mathbf{x}}^{k+1} = \mathbf{x}^{k+1} +
    \theta (\mathbf{x}^{k+1} - \mathbf{x}^k)
\end{cases}
\end{equation}
where $\theta \in [0, 1]$ ($\theta=1$ will be used in our work), $\tau$ and $\mu$ represent the step-lengths of the two sub-gradients. To ensure convergence $\tau \mu L^2 < 1 $ ($L^2 = ||\textbf{K}||_2^2 = \lambda_{max}(\mathbf{K}^H\mathbf{K})$ - i.e., the spectral radius of the linear operator.

\subsection{Inversion}
We start by analysing equation 2a and showing that it is possible to write it to be on the form of equation 4. Specifically, we write $\mathbf{K}=\nabla$, $\mathbf{z} = -\alpha \mathbf{p}^{k-1}$, $f=||\;||_{2,1}$, and $g = \frac{1}{2} ||\mathbf{d} - \mathbf{Gm}||_2^2 + \delta \sum_{j=1}^{N_c} \sum_{i=1}^{N_x N_z} V^{k-1}_{ji} (m_i - c_j)^2$. For simplicity, we rearrange some of the terms in $g$ such that we can write it as $g = \frac{1}{2} ||\hat{\mathbf{d}} - \hat{\mathbf{G}} \mathbf{m}||_2^2$:
\begin{equation}
\label{eq:augoperator}
\hat{\mathbf{V}} =
\begin{bmatrix}
diag \{\sqrt{\mathbf{V}_1^T}\} \\
... \\
diag \{\sqrt{\mathbf{V}_{N_c}^T}\}
\end{bmatrix}, \;
\hat{\mathbf{c}} =
\begin{bmatrix}
\sqrt{\mathbf{V}_1^T} c_0 \\
...\\
\sqrt{\mathbf{V}_{N_c}^T} c_{N_c}
\end{bmatrix}, \;
\hat{\mathbf{G}} =
\begin{bmatrix}
\mathbf{G} \\
\sqrt{2 \delta} \hat{\mathbf{V}}
\end{bmatrix}, \;
\hat{\mathbf{d}} =
\begin{bmatrix}
\mathbf{d} \\
\sqrt{2 \delta} \hat{\mathbf{c}}
\end{bmatrix}
\end{equation}

To conclude the proximal operator of $g$ is simply that of a quadratic functional which can be written as:
\begin{equation}
\label{eq:proxL2}
prox_{\tau g}(\mathbf{m}) =
        \left(\mathbf{I} + \tau \hat{\mathbf{G}}^T \hat{\mathbf{G}} \right)^{-1} \left( \mathbf{m} + \tau \hat{\mathbf{G}}^T \hat{\mathbf{d}}\right)
\end{equation}
Similarly, the proximal operator of $f$ is that of a $L_{2,1}$ norm - i.e. sum of the Euclidean norms of the columns of a matrix $\mathbf{Y}$ - which is:
\begin{equation}
\label{eq:proxL21}
prox_{\tau f}(\mathbf{Y}) = (prox_{\tau f}((\mathbf{Y}^T)_0)^T, ..., prox_{\tau f}((\mathbf{Y}^T)_N)^T)^T, \quad
prox_{\tau f}(\mathbf{y}) =        \left(1 - \frac{\sigma \tau}{max\{||\mathbf{y}||_2,
        \tau \}}\right) \mathbf{y}
\end{equation}

\subsection{Segmentation}
To start we rewrite equation 2c as follows:
\begin{equation}
\label{eq:segmentation}
\mathbf{v}^k = \underset{\mathbf{v} \in C} {\mathrm{argmin}} \quad
\mathbf{v}^T (\delta \mathbf{g} - \beta \mathbf{q}^{k-1}) + \beta \sum_{j=1}^{N_c} TV(\mathbf{V}^T_j)
\end{equation}

where $\mathbf{v} = Vec(\mathbf{V})$ and $\mathbf{q} = Vec(\mathbf{Q})$ are the vectorized version of $\mathbf{V}$ and  $\mathbf{Q}$, respectively, obtained by concatenating their rows into a vector. Moreover, we define the weighting vector $\mathbf{g} = ((\mathbf{m} - c_0)^2,..,(\mathbf{m} - c_{N_c})^2)^T$, where the superscript $2$ is used to indicate the squared difference between each element of the vector $\mathbf{m}$ and the chosen coefficient $c_j$. In its current form, equation 10 can be shown to be on the form of equation 4. This is achieved by writing $\mathbf{K}=\nabla$, $\mathbf{z} = \delta \mathbf{g} - \beta \mathbf{q}^{k-1}$, $f=||\;||_{2,1}$, and $g = i_C$ i.e., the indicator of the Simplex function. 

The algorithm described in equation \ref{eq:alg} is implemented here using the PyLops framework \cite{ravasi2020}.

\section*{Horizon extraction}
\begin{figure}
\centering
  \includegraphics[scale=0.45]{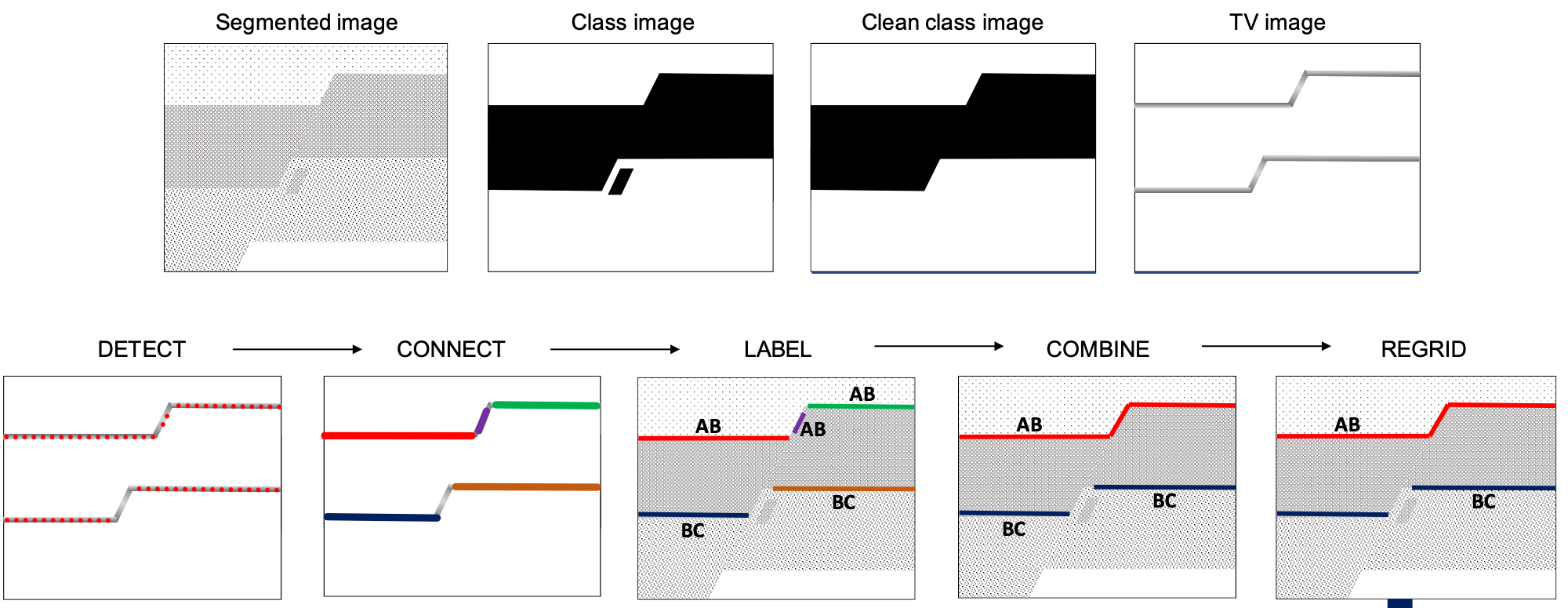}
  \caption{Workflow for the extraction of horizon lines from a single class. The segmented image is the starting point of the horizon extraction procedure from which a binary class image is computed prior to cleaning and performing a TV calculation to determine the edges of the class object. Neighbouring edge points are connected prior to being labelled based on surrounding classes. Finally similarly labelled lines are combined before being regridded to the full horizontal, spatial sampling of the model. The blue line on the X-axis of the regridding step illustrates where horizontal values are NaN.}
  \label{fig:hextract}
\end{figure}

Horizon extraction is performed per class as illustrated in Figure \ref{fig:hextract}. First, a binary class image is computed from the segmented image identifying the areas which are allocated to that class after the segmentation is complete. The binary class image undergoes two cleaning steps: a feature selection procedure that removes small closed foreground objects and morphological erosion which converts foreground pixels to the background pixel value when the background is dominant within a fixed window size. The terminology of foreground and background here represents the areas in the binary class image that have either been detected as that class or not.

Subsequently, the previously defined Total Variation is computed on the cleaned class image. Such a measure is frequently used for edge detection algorithms as it enhances the edges of objects isotropically. The TV computation identifies the boundaries of areas belonging to that class, i.e., the edges of the foreground objects, such that pixels with a high TV value are indicative of a boundary. Finally, a threshold is applied where every pixel above the threshold is determined as a potential horizon point.

Once detected, neighbouring horizon points are joined in a left-right, top-bottom fashion. Beginning at the top, most-left horizon point, $(x_0,y_0)$, neighbouring horizon points are identified within the limits $x_0<x<x_0+1$ and $y_0-1<y<y_0+1$. If there are multiple neighbouring points satisfying the TV threshold, the one with the highest TV value is selected as the next horizon point and the neighbour search begins again with the new points, $(x_1,y_1)$, searching the space, $x_1<x<x_1+1$ and $y_1-1<y<y_1+1$. The neighbour search continues until no points are detected in the neighbour space, concluding that line. At this point, a new neighbour search begins from the highest, most-left remaining horizon points to create another line. The colours and separated lines in the \textit{connect} image of Figure \ref{fig:hextract} represent the results of the neighbour search step. 
Once the neighbour search is exhausted, the lines are labelled using the segmented image to determine the class above and class below the lines, as illustrated in the \textit{label} image of Figure \ref{fig:hextract}. Three options exist for labelling the lines depending on the complexity of the geology and the homogeneity of the medium above and below the line. The labelling options are: only consider the class above, only consider the class below, or consider both classes above and below. 

Similar to the \textit{connect} step performed on a pixel level, a secondary combining procedure is performed on the labelled lines considering any lines that begin within a specified window from the ending of the previous line. Due to the smaller number of lines available to be joined, in comparison to the pixels in the \textit{connect}' step, a larger window can be considered allowing joining of points previously too far apart, as illustrated by the combining of the 3 $AB$ lines. For lines that are too far apart, such as $BC$, these will remain as separate lines to avoid any artefacts being generated in the regridding step.

Finally, the regridding step is performed for each identified horizon line. For continuous horizons, such as $AB$, the regridding procedure ensures a single depth value per horizontal location. On the other hand, when horizons present gaps (e.g., $BC$), the regridded horizon is further masked to avoid the creation of spurios connections. As a result, all of the identified horizons span the full lateral extent of the model.

Once the horizons have been labelled and extracted for each class, duplicate horizons exist where they have been identified as the top horizon from one class image and as the bottom horizon from another class image. The TV image values vary based on the class image therefore duplicate horizons are unlikely to be identical. For example in the case of the $BC$ horizon, the fault may have been detected when the class image represented the medium below the $BC$ horizon. Duplicate horizons arising from the above mentioned scenario are undesirable however there are other scenarios where horizons may be labelled the same based on their neighbouring classes where it is desirable to retain both, such as highly-layered settings. 

A horizon selection procedure is performed as follows:
\begin{enumerate}
    \item All horizons are initially set as ``to-be-kept''.
    \item The longest horizon, i.e. with the fewest NaN values, is identified and removed from the list of potential horizon lines.
    \item For each remaining horizon line, any shared points on the x-axis are compared and the average depth difference between the points is computed. 
    \item If the difference is less than a given threshold the horizon is considered a duplicate of the longest horizon and is removed from the list of potential horizon lines.
    \item If the difference is larger than the threshold, the horizon line remains on the potential horizon list and is still considered ``to-be-kept''.
    \item Steps 2-5 are repeated until there are no horizons left on the potential horizon list, i.e., until all horizons lines are either selected or rejected.
\end{enumerate}

The selected horizons, alongside the inverted and segmented models, ultimately represent the final outputs of our algorithm.

\section*{Examples}
In this section, we apply our methodology to two synthetic examples and a 2D line of the open-source Volve dataset.

\subsection*{Simple model}
The first example considers a very simple subsurface model composed of a number of horizontally stacked layers offset by a normal fault. Because of its simplicity, we will show that a single step of the proposed inversion scheme suffices to obtain a satisfactory segmentation of the model into different zones and for the tracking algorithm to extract all the horizons of interest. Nevertheless, despite its apparent simplicity, this model presents offsetted horizons which are historically challenging to track in seismic data.

\begin{figure}
\centering
  \includegraphics[height=8cm,keepaspectratio]{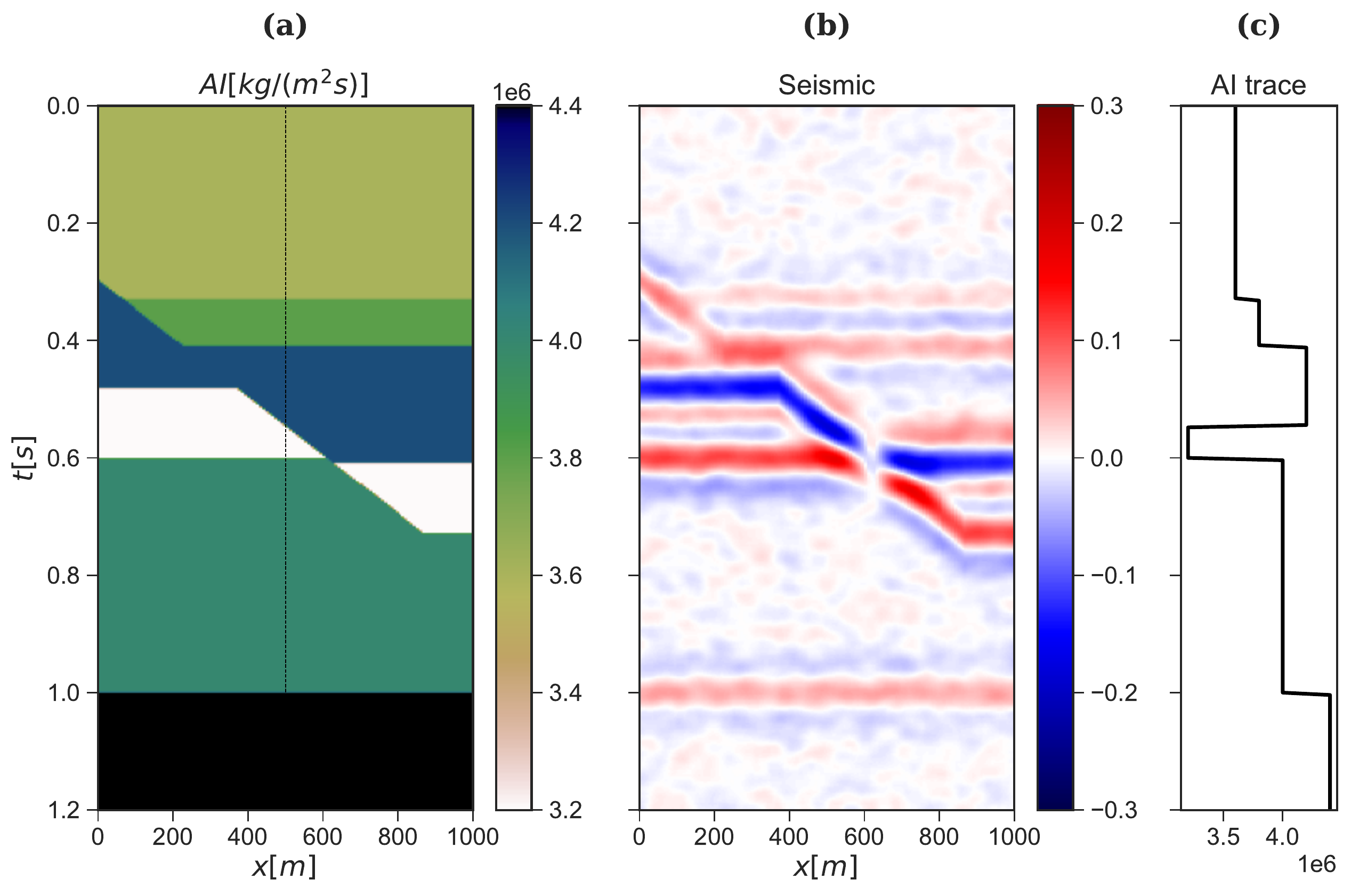}
  \caption{a) Acoustic impedance model, b) noisy seismic data, and c) AI profile at well location (shown by the dashed black line in panel a)}
  \label{fig:simplemod}
\end{figure}

Figure \ref{fig:simplemod}a displays the acoustic impedance model composed of six formations with different properties. For the sake of this exercise, we assume that a well is drilled in the middle of the model, such that all formations are sampled. The acoustic impedance profile is displayed in Figure \ref{fig:simplemod}c. The class vector $\mathbf{c}$ is therefore chosen to be a $6 \times 1$ vector whose elements are the different AI values in the vertical profile. The post-stack seismic data (Figure \ref{fig:simplemod}b) is modelled using the $\mathbf{G}$ operator discussed above and a 8Hz Ricker wavelet. Note that throughout the paper we define the vertical axis in terms of two-way traveltime for consistency with the field data example. Finally colored noise is added to the synthetic dataset: the noise model is created by smoothing a white noise realization along both the spatial and time axes. 

To begin with, the noisy dataset ($\mathbf{d_n} = \mathbf{G}\mathbf{m} + \mathbf{n}$) is inverted for the underlying acoustic impedance model ($\mathbf{m}$) using a number of approaches:
\begin{itemize}
	\item Least-squares regularized inversion, which minimizes the following cost function: $J_{L2} = ||\mathbf{d_n} - \mathbf{Gm}||_2^2 + \epsilon ||\mathbf{m}||_2^2 + \alpha ||\nabla^2 \mathbf{m}||_2^2$ where $\nabla^2$ is the Laplacian operator. The second term in the functional is the standard Thicknov regularization, whilst the third term enforces smoothness in the solution. The model $\mathbf{m}$ is here estimated by using an iterative solver for smooth, convex functionals such as LSQR using a smooth version of the model as starting guess. This results serves as a baseline as it represents the approach routinely used for post-stack inversion of seismic data.
	\item Linearized Alternating Direction Method of Multipliers (L-ADMM) with isotropic TV regularization \cite{Boyd}: $J_{L-ADMM} = \frac{1}{2} ||\mathbf{d_n} - \mathbf{Gm}||_2^2 + \alpha TV(\mathbf{m})$.
	\item Primal-dual algorithm with isotropic TV regularization \cite{Chambolle}:  $J_{PD} = J_{L-ADMM}$
\end{itemize}

\begin{figure*}[htb]
\centering
  \includegraphics[height=8cm,keepaspectratio]{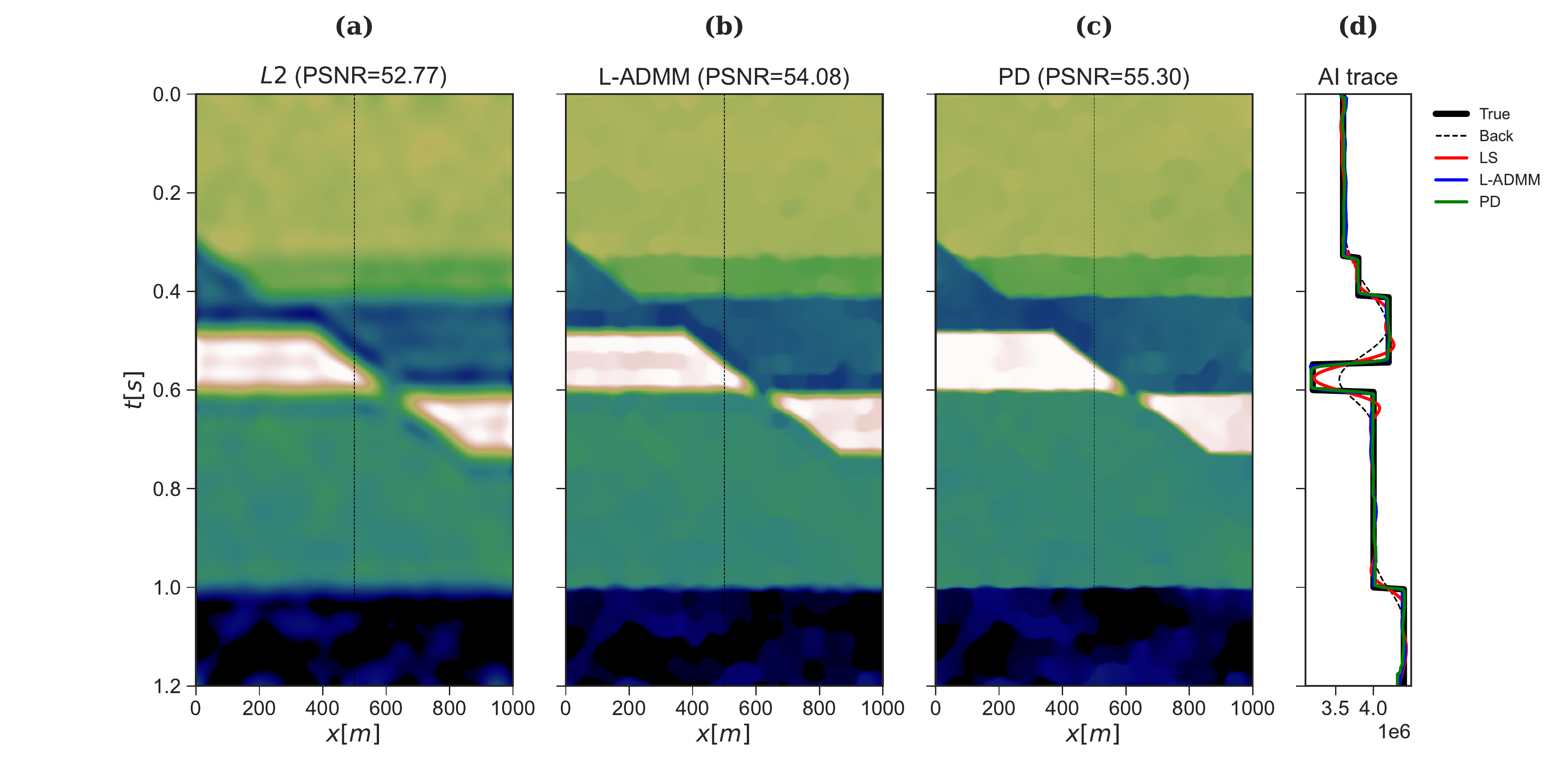}
  \caption{Inverted models via a) Least-squares regularized inversion, b) Anisotropic TV-regularized inversion Split-Bregman solver, and c) Isotropic TV-regularized inversion Primal-Dual solver. d) True, background, and inverted AI profiles at well location.}
  \label{fig:simpleinv}
\end{figure*}

In Figure \ref{fig:simpleinv} the different inversion results are compared. First, we can observe that whilst computationally cheaper than the other methods, standard least-squares inversion suffers from a major limitation: the retrieved model shows smooth transitions between different zones and therefore the sharp boundaries are not recovered. Whilst this is well-known limitation of such a method, it is particularly detrimental for our subsequent tasks of segmentation and horizon tracking as shown below. On the other hand both the L-ADMM and Primal-Dual algorithms are able to recover a model with sharp boundaries and an overall higher peak signal-to-noise ratio ($PSNR = 10 \log_{10}(N_xN_zmax(\hat{\mathbf{m}}) / ||\mathbf{m}-\hat{\mathbf{m}}||_2))$, where $\hat{\mathbf{m}}$ is the estimated model). Apart from producing an estimate with slightly higher PSNR, the latter solver possesses higher flexibility and the possibility to handle additional terms in the form of dot products as required in equation \ref{eq:alg}a of our joint inversion and segmentation algorithm that will be applied to the next example.

\begin{figure}
\centering
  \includegraphics[height=8cm,keepaspectratio]{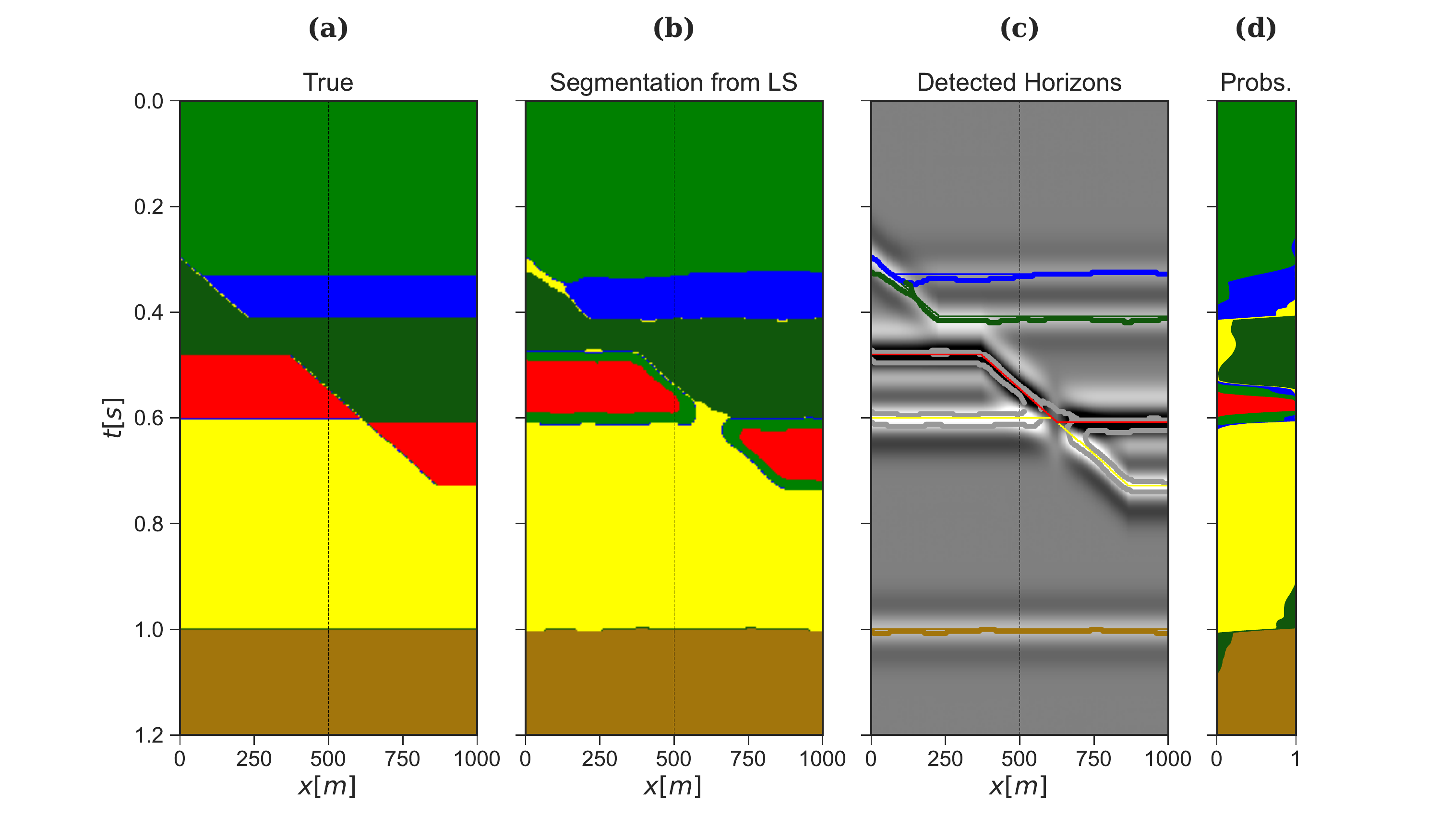}
  \caption{a) True and b) estimated segmentation from least-squares inverted model, respectively. c) Tracked horizons overlaid to the noise free seismic dataset (thin lines: true horizons, thick lines: tracked horizons. d) Probabilities for each class at well location (i.e., a single column of $\mathbf{V}$ at the index in the middle of the model.}
  \label{fig:simplesegl2}
\end{figure}

\begin{figure}
\centering
  \includegraphics[height=8cm,keepaspectratio]{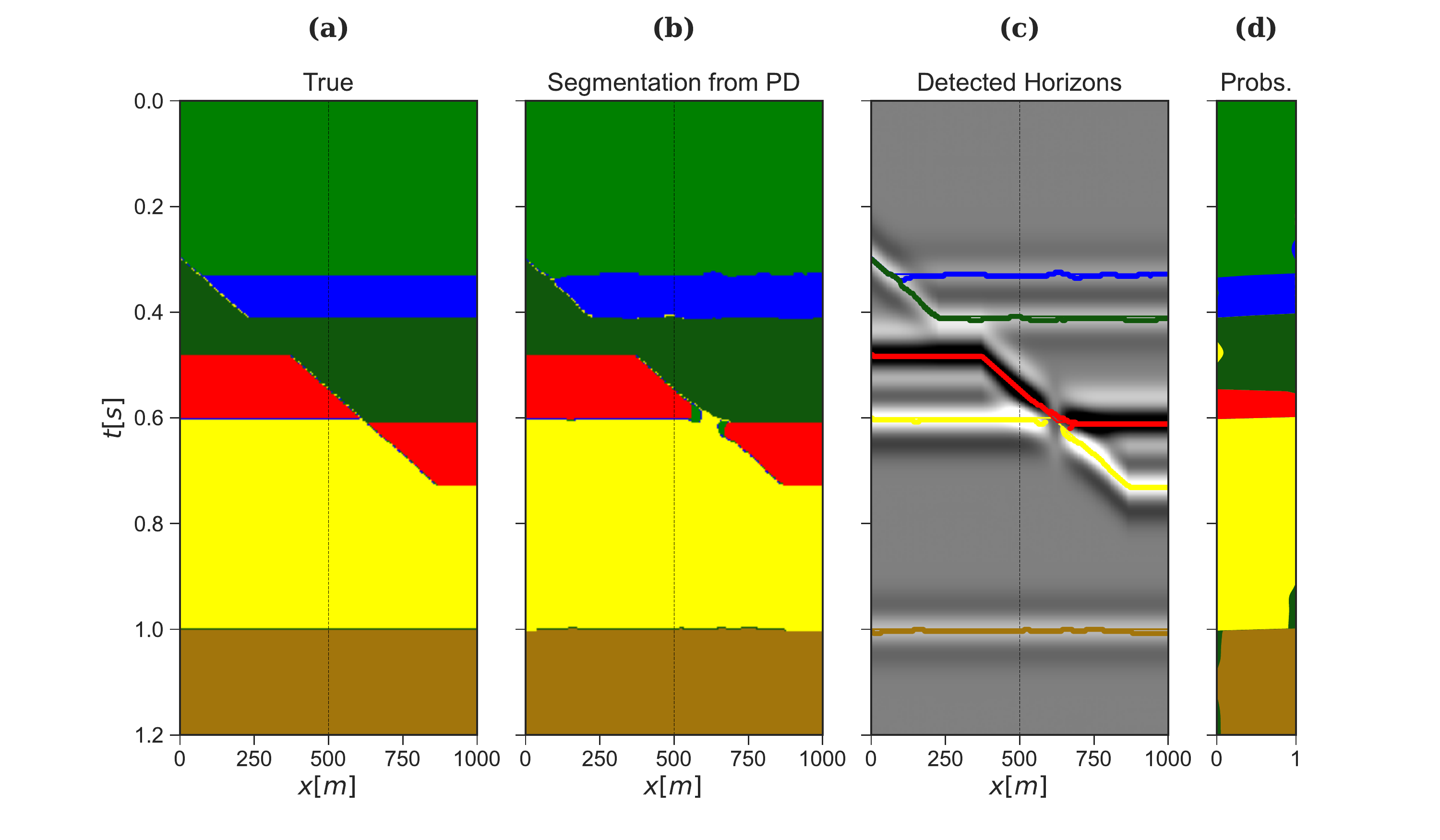}
  \caption{Same as Figure 3 with the Primal-Dual model used as input for segmentation.}
  \label{fig:simplesegpd}
\end{figure}

The models obtained by minimizing the first and third functionals are now used as input to the segmentation step. In mathematical terms, this step is equivalent to solving a simplified version of equation \ref{eq:alg}c which only contains the TV norm instead of the Bregman TV norm - in other words where $\mathbf{Q}$ is set to zero. Figures \ref{fig:simplesegl2} and \ref{fig:simplesegpd} show the results of the segmentation for the model produced by least-squares and Primal-Dual inversions, respectively. In both cases, panel a shows the true segmented model, where each image pixel is assigned the label of its class. In panel b the binarised version of the estimated matrix $\mathbf{V}$ is displayed; \textit{binarised} means that the index of the highest value of each column $\mathbf{V}$ is selected. The segmentation result from the least-squares model presents an evident problem inherited from the smooth transitions across interfaces observed in the inverted model; green rings appear all around the red class due to the fact that the model parameter of the green class lies in between that of the red and yellow classes. This becomes even more evident when observing the probabilities of each class in the vertical pillar in the middle of the model (Figure \ref{fig:simplesegl2}d). When attempting to track horizons, the presence of such rings affect the overall quality of the estimated horizons. Moreover, some of the erroneously tracked horizons do not conform with the layering knowledge provided by the input well log.

On the other hand, the segmentation results obtained when using the model estimated via the Primal-Dual solver are much more accurate. Note also that the probabilities estimated by the same number of iterations of the Primal-Dual algorithm used for segmentation present more sharp transitions between the different zones. This greatly benefits the subsequent step of horizon tracking which returns the five main horizons with a very high degree of precision. Overall, we can conclude that the process of identifying seismic horizons by means of tracking of the edges of closed shaped polygons seems to be less cumbersome and more robust than the problem of tracking a horizon in a highly-oscillatory multi-dimensional signal such as the seismic data itself.

\subsection*{Hess model}
\begin{figure}
\centering
  \includegraphics[height=5.5cm,keepaspectratio]{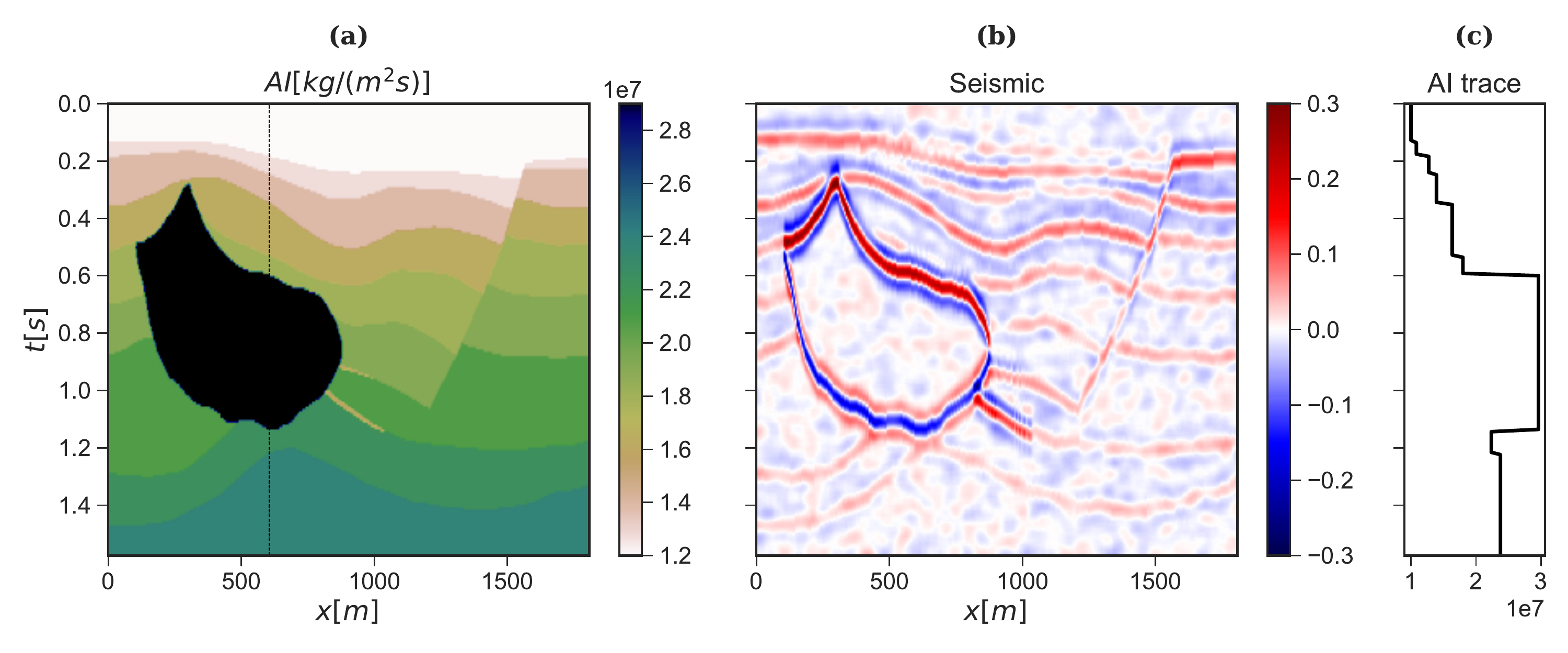}
  \caption{a) Acoustic impedance model, b) noisy seismic data, and c) AI profile at well location (shown by the dashed black line in panel a)}
  \label{fig:hessmod}
\end{figure}

Our second example is based on a modified version of the P-wave velocity of the SEG Hess VTI model (Figure \ref{fig:hessmod}a). This model is chosen for two reasons: first, we want to investigate the ability of our method to deal with stratigraphies that are interrupted by intrusions such as the salt body present in this model. Second, we are interested to assess the importance of our joint inversion-segmentation approach in the presence of very strong impedance contrasts such as those originated by the salt body.

To begin with, we consider once again the case where we have complete knowledge of the different zones in the model and their corresponding acoustic impedance value - this is the case where a well has been drilled through the salt body as shown in Figure 5c.  Similar to the previous example, the post-stack seismic data (Figure 5b) is modelled using the $\mathbf{G}$ operator with a 8Hz Ricker wavelet.

\begin{figure}
\centering
  \includegraphics[height=5cm,keepaspectratio]{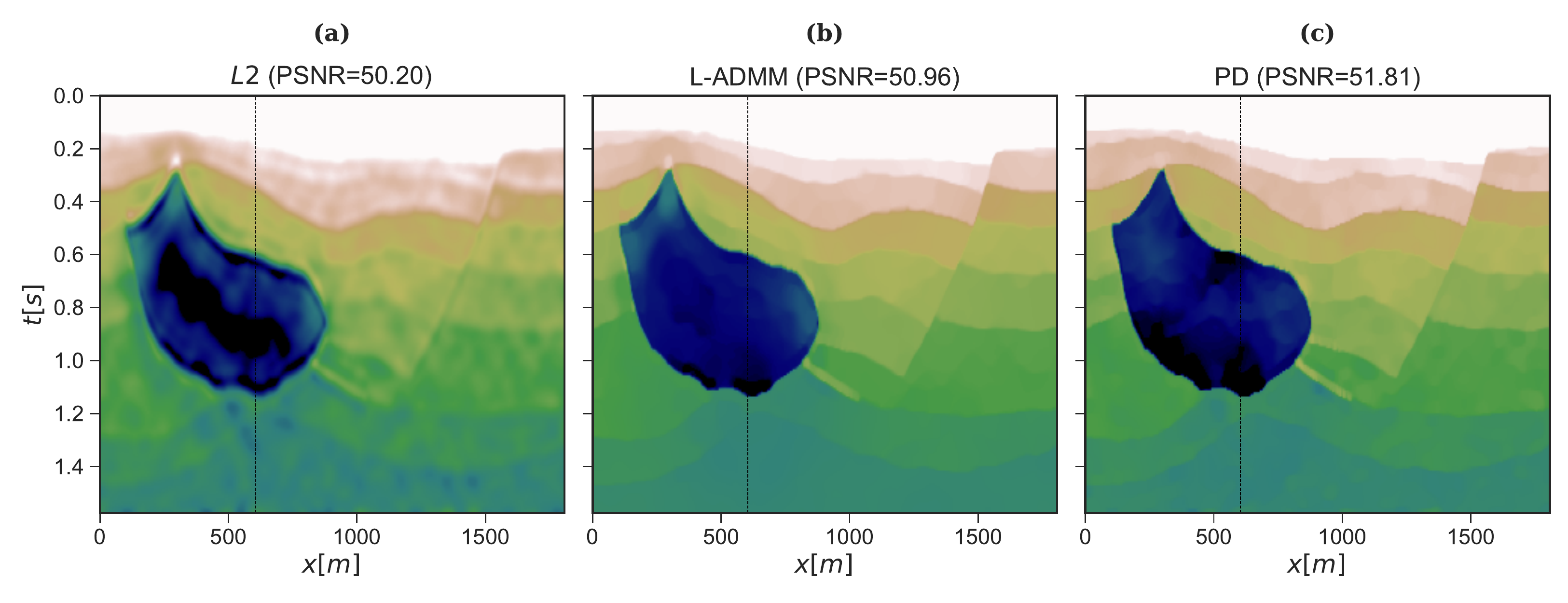}
  \caption{Inverted models via a) Least-squares regularized inversion, b) Anisotropic TV-regularized inversion Split-Bregman solver, and c) Isotropic TV-regularized inversion Primal-Dual solver. d) True, background, and inverted AI profiles at well location.}
  \label{fig:hessinv}
\end{figure}

Figure \ref{fig:hessinv} shows the estimated model for three different solvers: LSQR (used for the least-squares regularized functional), L-ADMM and PD (used the TV regularized function). Once again, the inversion of noisy data benefits from use of the isotropic TV norm which regularizes the solution and produces sharp discontinuities at the edges of the model. Nevertheless, a common feature of the three inverted models is represented by the inaccurate estimate of the acoustic impedance of the salt body which is underestimated by all of the tested algorithms. This is not surprising as the background model is fairly far from the true solution, especially towards the edges of the salt body, and the frequency content of the seismic data does not allow to recover such a gap even when a strong regularization such as the isotropic TV norm is included in the functional to optimize.

\begin{figure}
\centering
  \includegraphics[height=5.5cm,keepaspectratio]{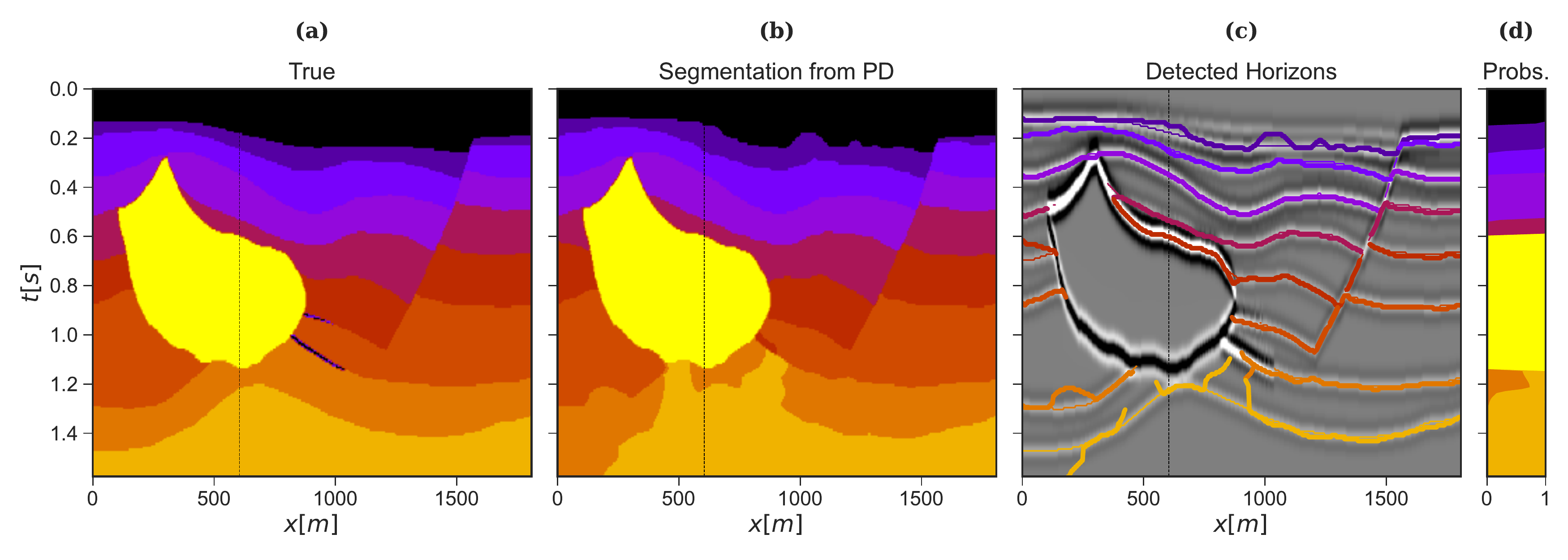}
  \caption{a) True and b) estimated segmentation from Primal-Dual model, respectively. c) Tracked horizons overlaid to the noise free seismic dataset (thin lines: true horizons, thick lines: tracked horizons. d) Probabilities for each class at well location (i.e., a single column of $\mathbf{V}$ at the index in the middle of the model.}
  \label{fig:hessseg}
\end{figure}

Despite of the observed inaccuracies in the retrieved acoustic impedance model, segmentation and horizon tracking still produce satisfactory results when applied to the Primal-Dual model (Figure \ref{fig:hessinv}). Note that most of the zones in the original model (Figure \ref{fig:hessseg}a) have been segmented precisely (Figure \ref{fig:hessseg}b) with only the shallowest and two deepest zones being quite noisy due to the weak impedance contrast affected by noise in the data (Figure \ref{fig:hessmod}b). A similar conclusion can be drawn for the outcome of the horizon tracking step (Figure \ref{fig:hessseg}c) where most of the horizons overlay almost perfectly with their true counterpart apart from those related to the poorly recovered zones. 

\begin{figure}
\centering
  \includegraphics[height=5.5cm,keepaspectratio]{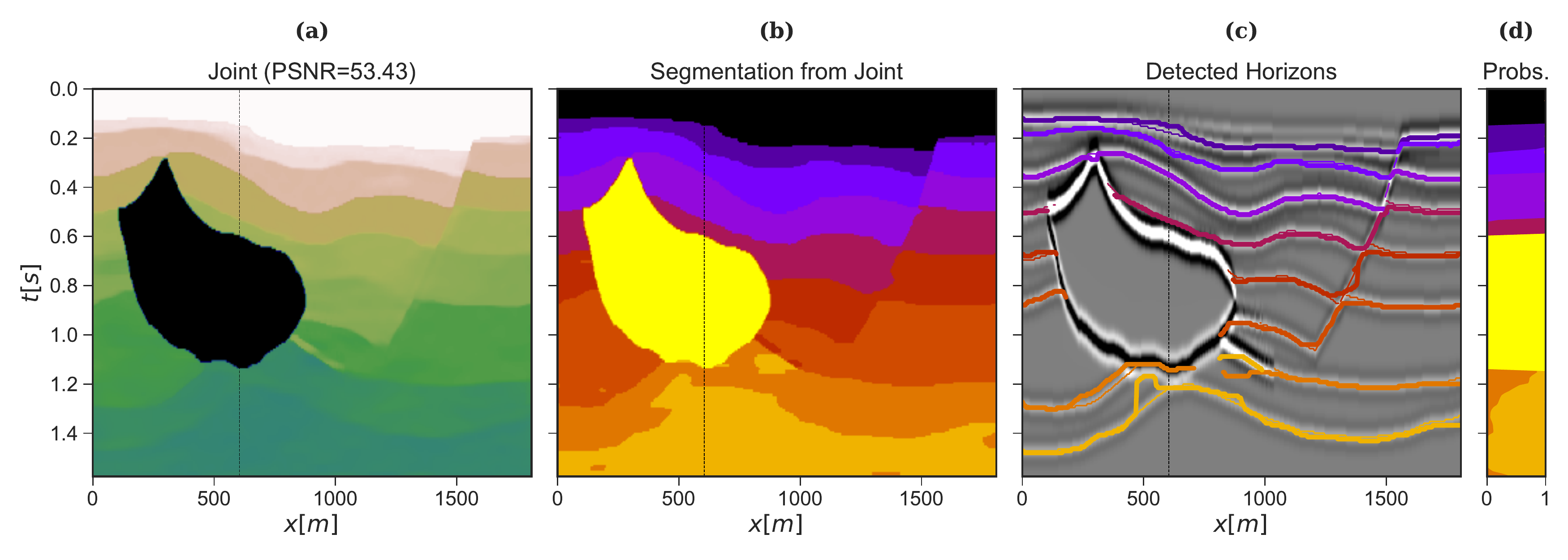}
  \caption{a) Model and b) segmentation from joint inversion scheme. c) Tracked horizons overlaid to the noise free seismic dataset (thin lines: true horizons, thick lines: tracked horizons. d) Probabilities for each class at well location (i.e., a single column of $\mathbf{V}$ at the index in the middle of the model.}
  \label{fig:hessjoint}
\end{figure}

\begin{figure}
\centering
  \includegraphics[height=6.5cm,keepaspectratio]{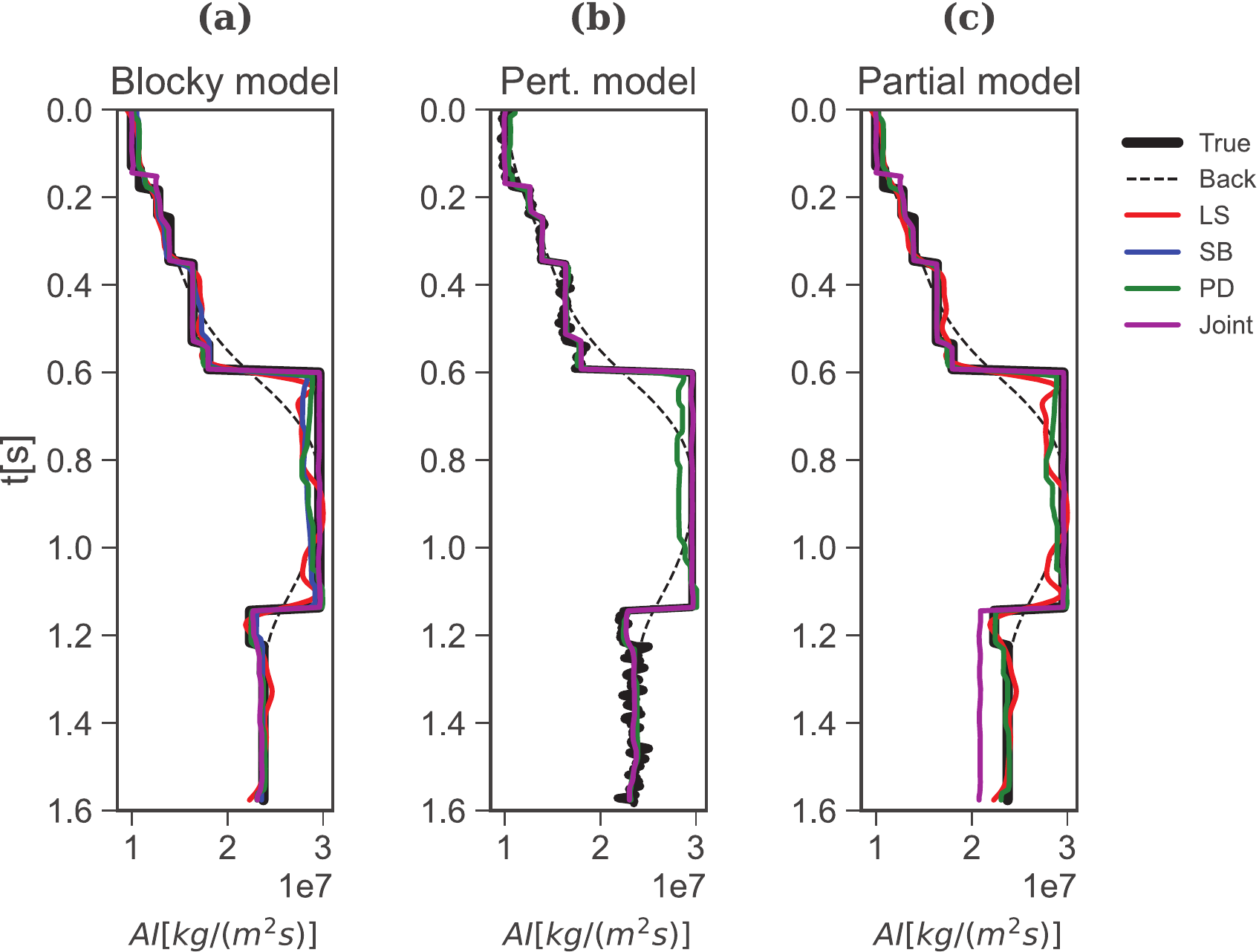}
  \caption{Acoustic impedance estimates for the different inversion algorithms at well location for Hess dataset. a) Blocky model in Figure \ref{fig:hessmod}a, b) Model with small scale perturbations in Figure \ref{fig:hesspert}a, and c) Model with short well in Figure \ref{fig:hesspartial}a.}
  \label{fig:hesstraces}
\end{figure}

The joint inversion-segmentation algorithm proposed in the previous section is now employed to evaluate whether further constraining repeated steps of inversion with previously segmented zones could further improve the results in Figures \ref{fig:hessinv} and \ref{fig:hessseg}. The retrieved model in Figure \ref{fig:hessjoint}a after 4 outer iterations of the joint algorithm does indeed more closely resemble the true model. This is especially the case for the salt body, whose absolute value is not underestimated in this case (Figure \ref{fig:hesstraces}a), as well as the overall continuity of the different layers. This is the consequence of the fact that the first step of segmentation is used as soft constraint to the second step of inversion (last term in equation \ref{eq:alg}a), whose output is in turn used to drive the second step of segmentation (first term in equation \ref{eq:alg}c) and so on and so forth. The outcome of the last segmentation step (Figure \ref{fig:hessjoint}b) is also clearly more accurate than the one produced by directly using the Primal-Dual model: we observe a better continuity of the top interface of the shallowest zone as well as closer resemblance to the true classes in the deeper parts of the model. Similarly, the tracked horizons are of overall higher quality especially on the left of the salt body and in the deeper part of the model.

\begin{figure}
\centering
  \includegraphics[height=10cm,keepaspectratio]{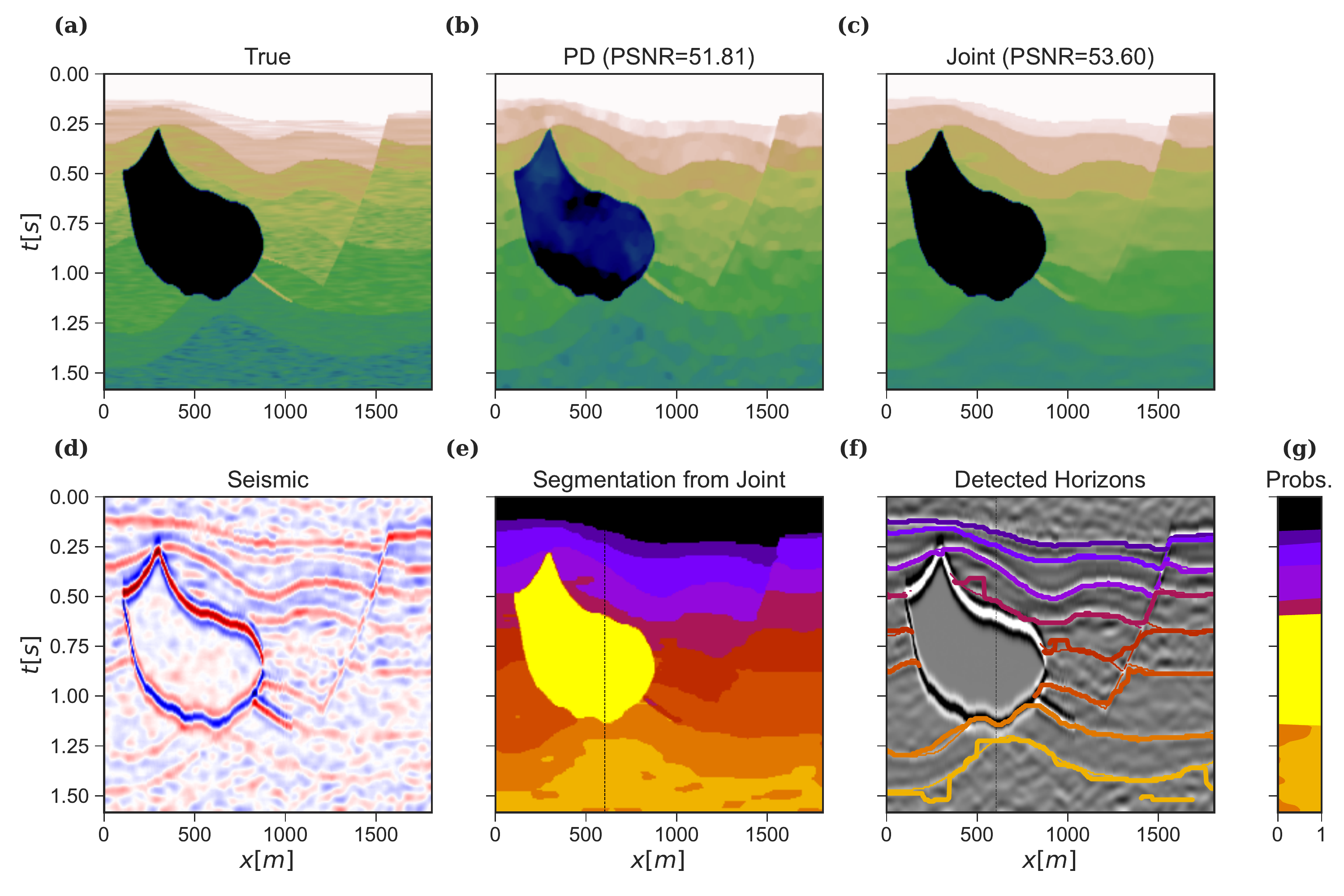}
  \caption{Results for model with small scale variations. a) True model, b) model from Primal-dual inversion, c) model from joint inversion scheme, d) noisy data, e) segmentation from the joint inversion scheme, f) tracked horizons overlain on the noise-free seismic dataset (note that noisy features are due to small scale AI perturbations in the model), and g) probabilities for each class at well location.}
  \label{fig:hesspert}
\end{figure}

A second scenario is now evaluated. The acoustic impedance model in Figure \ref{fig:hessmod}a is modified by adding small scale fluctuations to the macro model composed of a limited number of acoustic impedance values. Such small scale variations are clearly visible in the log data (Figure \ref{fig:hesstraces}b) and further complicate the overall seismic response (Figure \ref{fig:hesspert}e). The definition of our classes $\textbf{c}$ is therefore made by blocking the acoustic impedance log. We clearly do not expect our inversion to be able to recover such features given that they are outside of the bandwidth of the signal: on the other hand, this examples serves the purpose of testing the sensitivity of our algorithm to more complex and realistic models of the subsurface. Panels c, e, and f in Figure \ref{fig:hesspert} show that the different steps of our joint inversion are robust to small scale variations in the model and the results are overall very similar to those of the ideal case with a well defined macro model. Whilst all the horizons are successfully tracked in the segmented model, some spurious events are also reconstructed as part of the deeper horizon. This is the direct consequence of small blobs in the deeper layer of the segmented model (\ref{fig:hesspert}e) which lead our tracking algorithm to identify those lines as part of the same horizon group as the deepest horizon because they border same pair of classes as the larger horizon does.

\begin{figure}
\centering
  \includegraphics[height=5.5cm,keepaspectratio]{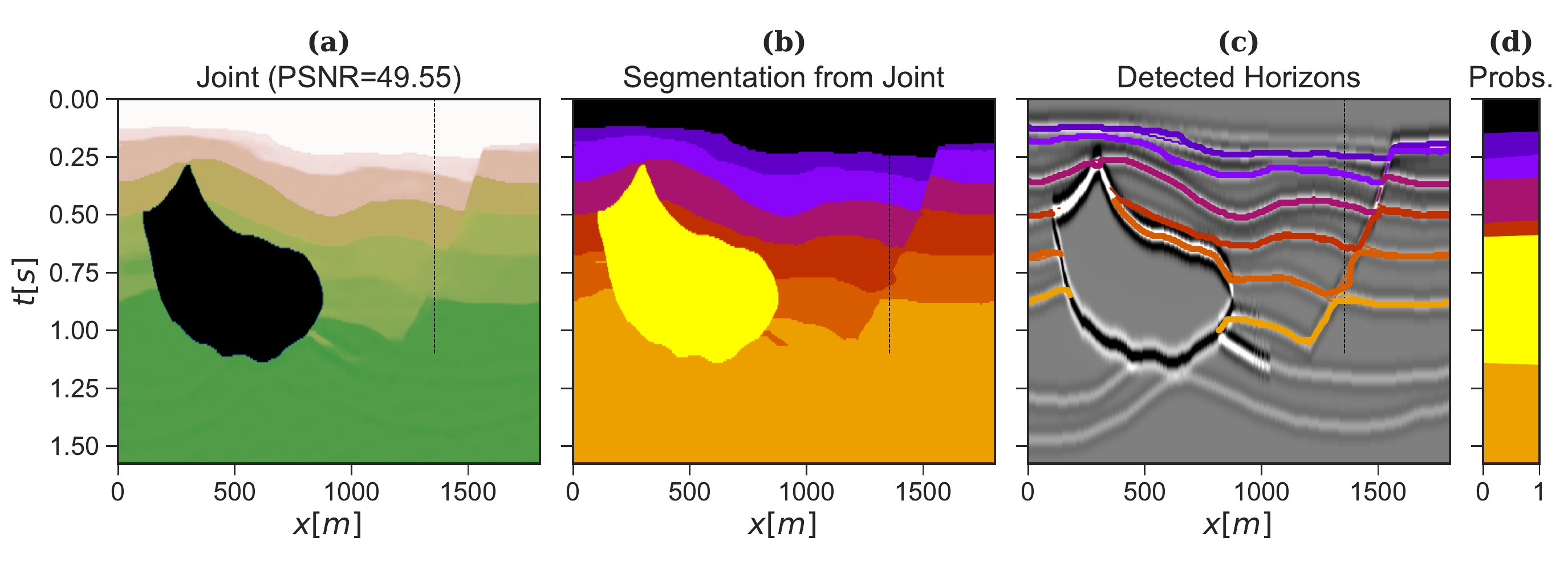}
  \caption{Results for model with short well. a) Inverted model from joint inversion scheme, b) segmentation from the joint inversion scheme, and f) tracked horizons overlain on the noise-free seismic dataset.}
  \label{fig:hesspartial}
\end{figure}

Finally, we investigate the robustness of our algorithm with respect to partial knowledge of the different classes. As shown in Figure \ref{fig:hesspartial} the vertical well is now assumed to penetrate only some of the layers in the model: as a consequence, the classes vector $\textbf{c}$ only contains seven elements to which we add an eighth based on the assumption that we are aware of the presence of the salt body and have a good estimate of its acoustic impedance. Whilst our algorithm fails to track the two deepest interfaces, both the inversion and horizon tracking of the shallow subsurface is still satisfactory and not affected by the lack of knowledge of the deeper part of the model. On the other hand, as shown in Figures \ref{fig:hesspartial}a and Figures \ref{fig:hesstraces}c, the acoustic impedance estimates in the deeper section are underestimated as a consequence of the fact that our joint algorithm tends to drive acoustic impedance values at each location closer to those of the class that has been selected in the previous iteration.

\subsection{Volve dataset}
Volve is an oil field located in the central part of the North Sea, five kilometres north of the Sleipner Øst field. Volve produced oil from sandstone of Middle Jurassic age in the Hugin Formation, with the main reservoir located at a depth of approximately 2,700-3,100 metres. The field was shut down in 2016, with the facility removed in 2018, and all historical subsurface and production data made available by Equinor in June 2018.

\begin{figure}
\centering
  \includegraphics[height=5cm,keepaspectratio]{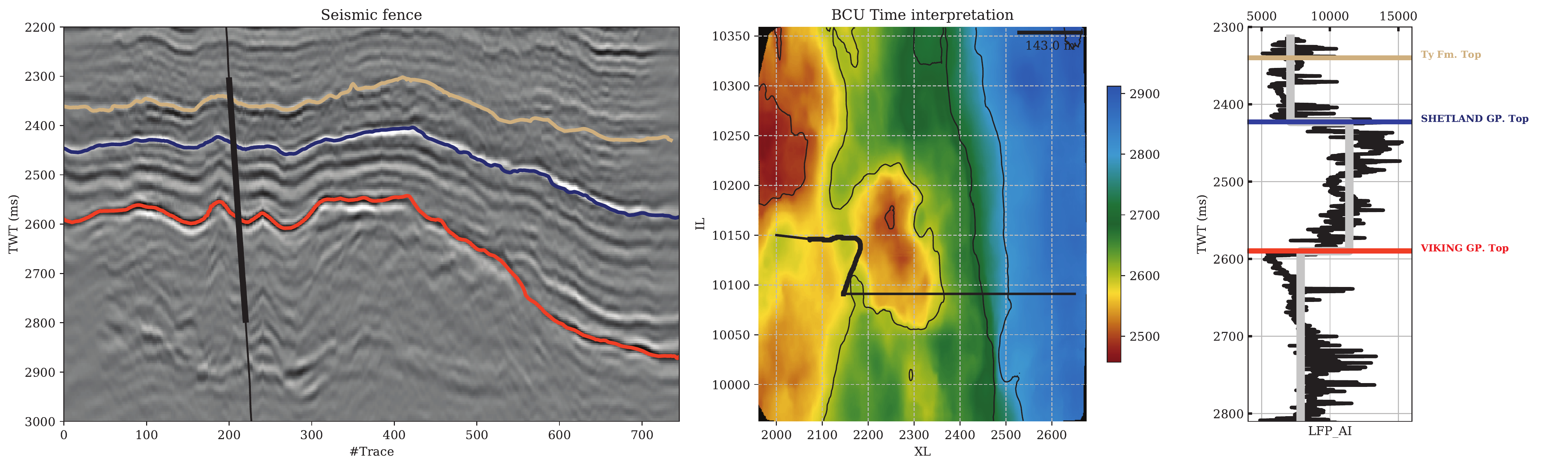}
  \caption{a) Seismic section along the NO/15-9 19 BT2 well with Top Ty (brown line), Top Shetland (blue line) and BCU (red line) horizons. b) BCU time horizon with well trajectory and fence along which seismic is extracted and displayed in the previous panel. c) Acoustic impedance well log and well markers for the 3 formations of interest. All data displayed here is taken from the official Volve dataset.}
  \label{fig:volvedata}
\end{figure}

In this section, our joint inversion and segmentation algorithm is applied to a 2d section of the PSDM full stack dataset from the ST10010ZC11 survey (Figure \ref{fig:volvedata}a). Since interpretation has been carried out in time, the time version of this dataset is used in our example. The section is extracted along the NO/15-9 19 BT2 well and further extended to the East of the well as shown on top of a time map of the BCU time surface (Figure \ref{fig:volvedata}b). Figure \ref{fig:volvedata}c shows the acoustic impedance log converted into two-way traveltime (TWT) using the available checkshot profile. The 3 key interpreted horizons are overlain on the seismic section in Figure \ref{fig:volvedata}a and their corresponding well markers are shown in Figure \ref{fig:volvedata}c. 

\begin{figure}
\centering
  \includegraphics[height=5.5cm,keepaspectratio]{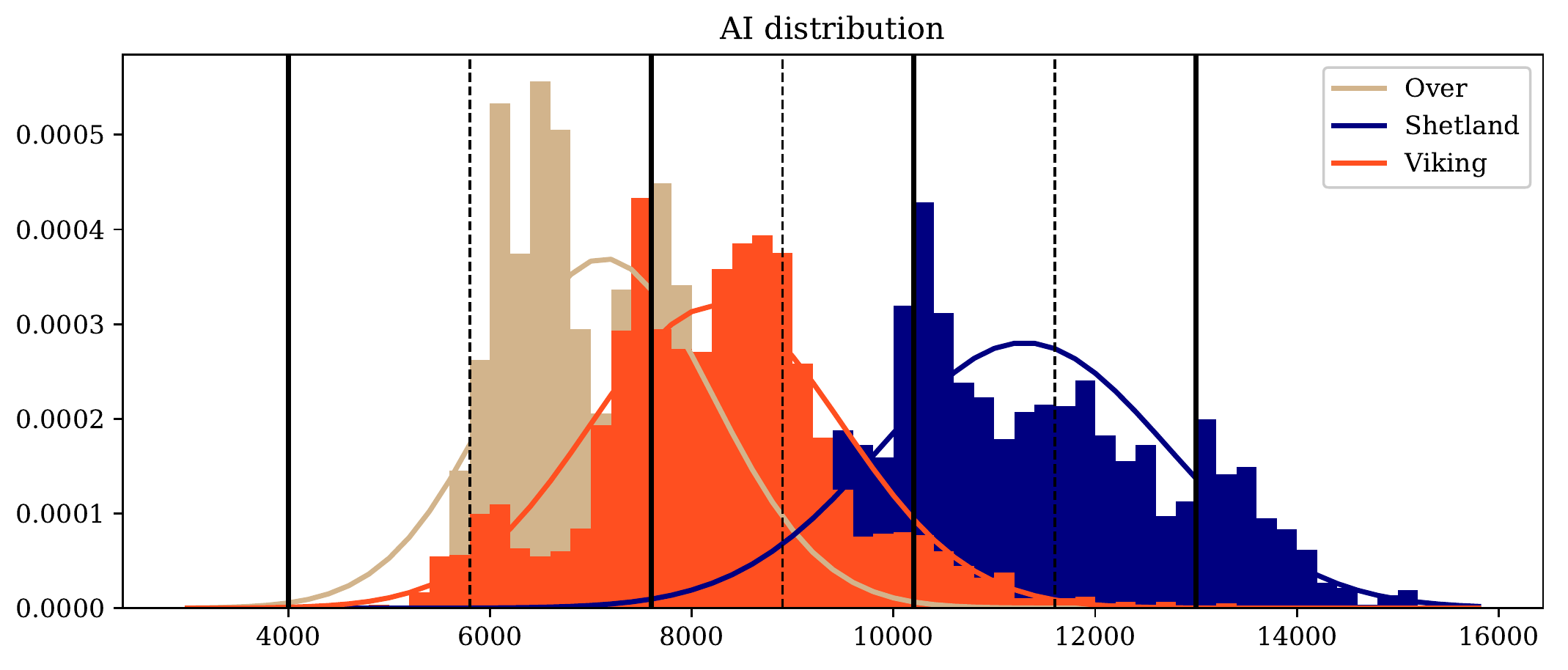}
  \caption{Histogram of the acoustic impedance values at NO/15-9 19 BT2 well divided in zones as defined in Figure \ref{fig:volvedata}c. Solid black vertical lines refer to the classes chosen for the segmentation algorithm whilst dashed black vertical lines represent the separation between the classes.}
  \label{fig:volvehist}
\end{figure}

To begin with, the acoustic impedance profile is used to identify classes for segmentation step. Three different zones are chosen, separated by the Shetland and Viking tops and the histograms of their acoustic impedance values plotted in Figure \ref{fig:volvehist}. However, given the large overlap in the histograms from the overburden class and the Viking class, they are aggregated into a single class. A second class is defined to include most of the values in the Shetland formation and two other classes are defined to capture the low and high acoustic impedance values observed in the well log. Ultimately, 4 different classes ($\textbf{c}=[4000, 7600, 10200, 13000]$) are used as input to our segmentation algorithm. Note that our ability to successfully separate different formations is highly dependant on the choice of these classes. To accurately fine tune such a step, it may be useful to perform a first independent inversion step and apply a pixel-wise segmentation based on the closeness of each pixel value to the classes values. This is equivalent to solving the segmentation problem in equation \ref{eq:alg}c without the TV-regularization term ($\beta=0$) leading to a very noisy segmentation result. Nevertheless, by closely inspecting this segmentation the classes boundaries can be further optimized prior to running the entire joint inversion and segmentation scheme.

\begin{figure}
\centering
  \includegraphics[height=8cm,keepaspectratio]{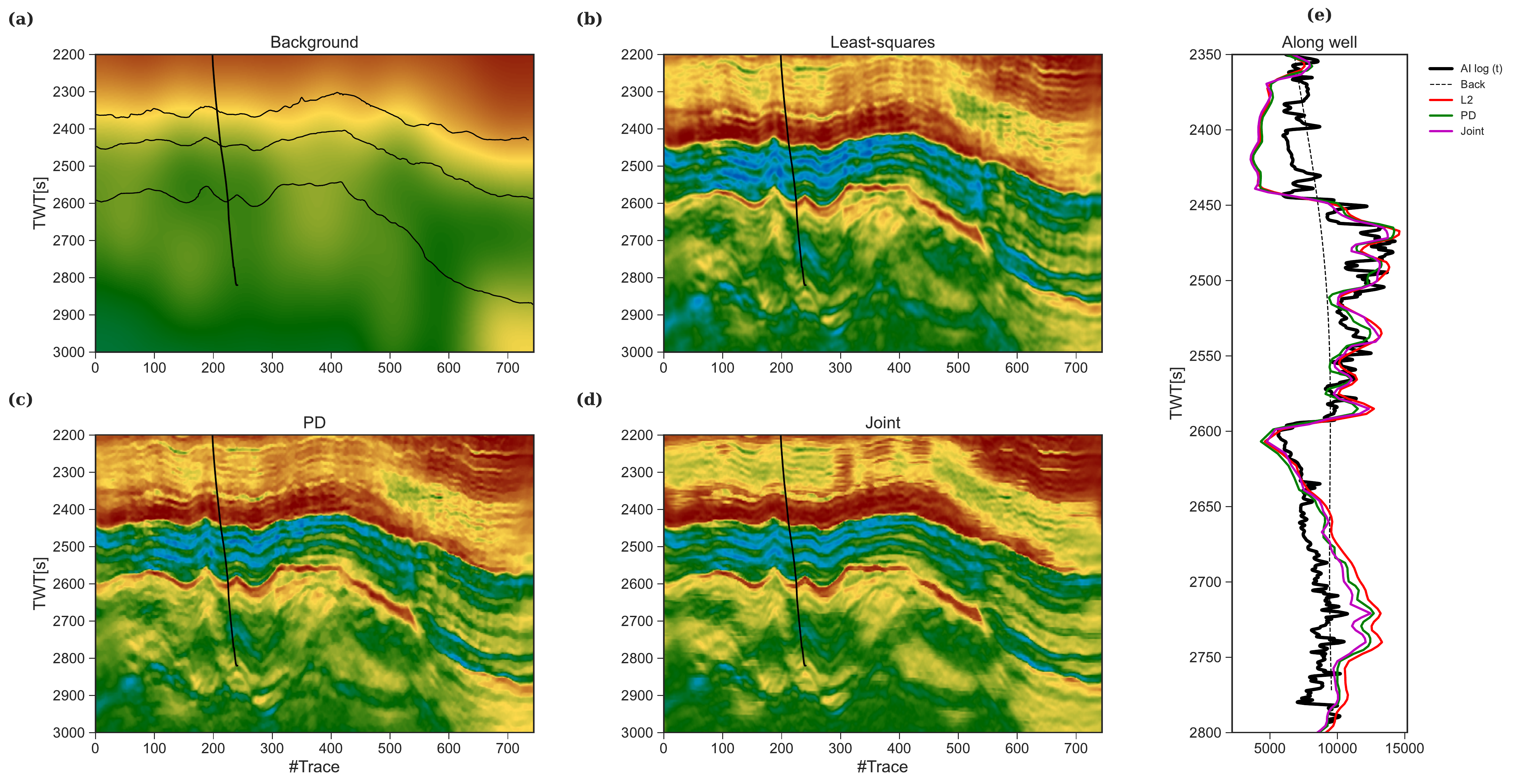}
  \caption{Acoustic impedance models for the Volve dataset: a) background, b) regularized least-squares, c) primal-dual, and d) joint inversion. e) AI values at well location for the different inversions.}
  \label{fig:volveinv}
\end{figure}

To be able to invert the seismic data for an acoustic impedance model, a background model is built from the root-mean-square (RMS) velocity model. RMS velocities are first converted into interval velocities and subsequently calibrated with the acoustic impedance log of the NO/15-9 19 BT2 well. More specifically, the interval velocity model is extracted along the well trajectory and used to `predict' a smoothed version of the acoustic log: prediction is achieved via linear regression and the regression coefficients are further used to convert the entire velocity model into a background acoustic impedance model (Figure \ref{fig:volveinv}a). Note that other approaches to construct the low-frequency model can be equivalently used in cases where a larger coverage of vertical wells with the required set of logs is available. Estimated acoustic impedance models for three different optimization problems are shown in the other panels of Figure \ref{fig:volveinv}: from left to right, spatially regularized least-squares inversion, TV regularized inversion, and our joint inversion-segmentation. As already observed in the other examples, TV regularized inversion improves on the blockiness of the model compared to least-squares inversion and produces sharper transitions between different formations; this is especially the case for areas with low and high acoustic impedance values which are generally under predicted due to the lack of low frequency information in the data. After two outer iterations of the algorithm in equation \ref{eq:alg}, this behaviour is even more visible as a consequence of the second regularization term in equation \ref{eq:joint}, which drives acoustic impedance values closer to that of the class they in which they belong to. Moreover, whilst the other two inversion results show some vertical striping in an area of poorer data quality (white arrows), the joint inversion manages to improve the lateral continuity of the model without compromising on the sharpness of the vertical transitions. Finally, an overall good match is observed with the acoustic impedance log along the well trajectory as shown in Figure \ref{fig:volveinv}e.

Figure \ref{fig:volveseg} displays the segmented model at the end of the joint inversion scheme alongside with the probabilities of each class along the well trajectory. White lines represent the different horizons that have been extracted from the different class probabilities. Due to the complexity of the model and the fact that some pairs of classes repeat at different depths, the horizon tracking algorithm is run in this case only considering the class above during the labelling and combination steps. The tracked horizons are also shown in Figure \ref{fig:volveseg}c alongside the manually interpreted horizons (thin black lines) on top of the input seismic data. A number of interesting observations can be made with respect to the ability of our algorithm to interpret the key horizons in this data: first, the Top Shetland formation is successfully tracked and the resulting horizon is very similar to the manually interpreted one. On the other hand, our interpretation of Top Ty diverges from the manual interpretation towards the right side of the model; a similar behaviour is also observed for the horizon above (red line in Figure \ref{fig:volveseg}c). By looking at the inverted acoustic impedance (Figure \ref{fig:volvesegzoom}b), we can observe a thinning of the Ty formation which is consistent with the tracked horizons. Whilst the quality of the seismic data in this area does not allow to determine whether our interpreted horizons are correct, this result highlights the direct connection between the tracking algorithm and the inversion step. Future work will investigate this behaviour by both inverting the entire 3d dataset and looking at its consistency in the perpendicular direction as well as looking at the pre-stack data in this area. Finally, our algorithm is only able to track part of the BCU horizon, failing to do so when the thin low acoustic impedance layer is eroded out on the right side of the model. Overall, the results in this section confirm the validity of our algorithm and its usefulness in jointly solving a number of tasks and keeping consistency among the different results: in other words, whilst the acoustic impedance model alone may be useful to condition facies and property modelling, the segmentation model may complement it in defining different areas of influence which may be conditioned differently. 

\begin{figure}
\centering
  \includegraphics[height=6cm,keepaspectratio]{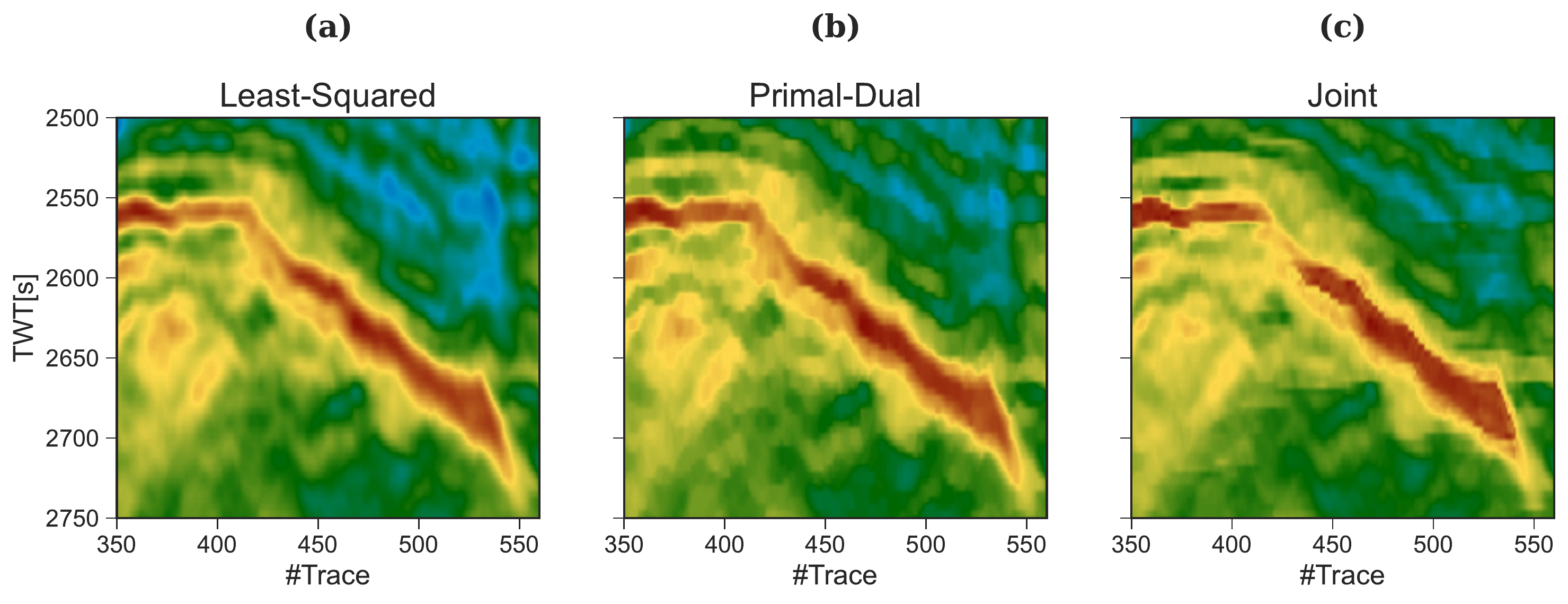}
  \caption{Zoomed sections of acoustic impedance inversion for a) least-squares solver, b) Primal-dual solver and c) Joint inversion.}
  \label{fig:volveinvzoom}
\end{figure}

\begin{figure}
\centering
  \includegraphics[height=14cm,keepaspectratio]{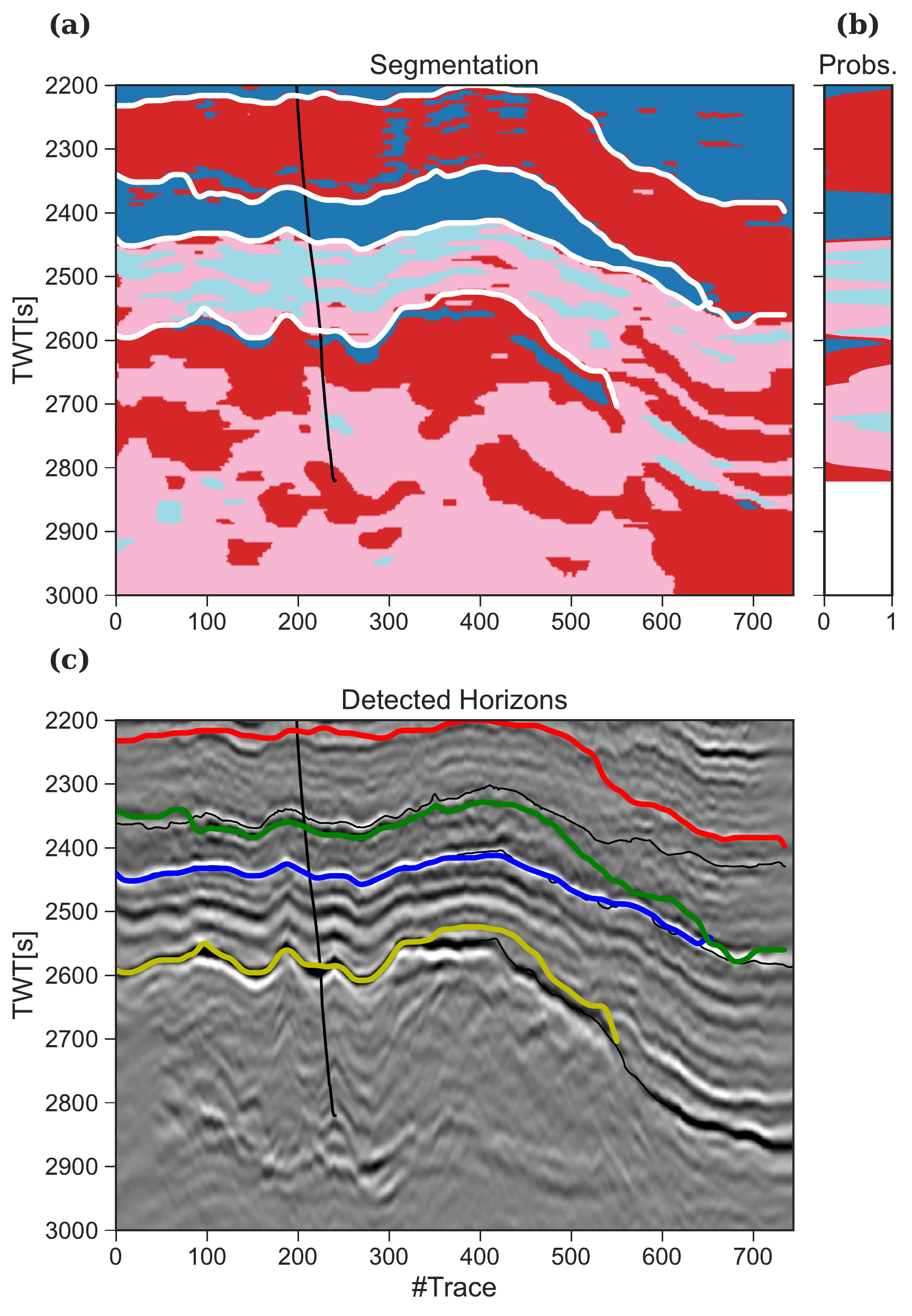}
  \caption{a) Segmentation model with tracked horizons (white lines), b) seismic data with tracked horizons, and c) probabilities for each class at well location}
  \label{fig:volveseg}
\end{figure}

\begin{figure}
\centering
  \includegraphics[height=10cm,keepaspectratio]{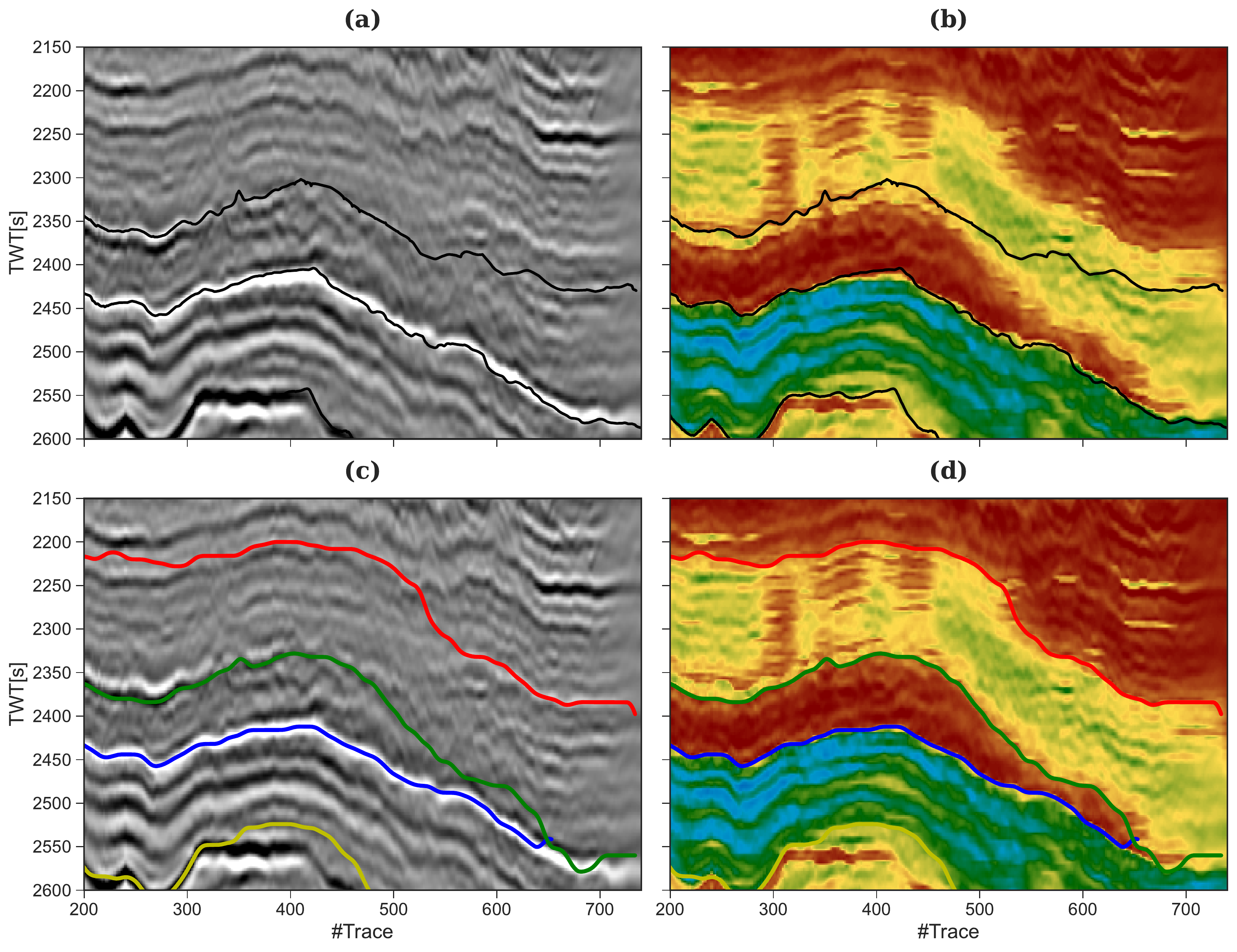}
  \caption{Zoomed sections of (a-c) seismic data and (b-d) acoustic impedance inversion with interpreted horizons (top: original interpretation, bottom: our interpretation).}
  \label{fig:volvesegzoom}
\end{figure}

\section*{Discussion}
\subsection{Computational considerations}
The computational cost associated with the Primal-Dual algorithm used to optimize equations \ref{eq:alg}a and \ref{eq:alg}c is highly dependant on that of the proximal operators ($prox_f$ and $prox_g$) involved in each iteration. As already shown in equation \ref{eq:proxL21}, the proximal operator of the $L_{2,1}$ requires only point-wise evaluations for each element of the input vector; this is therefore very fast to compute even for very large two or three dimensional models. 

On the other hand, the evaluation of the proximal operator for the quadratic functional $\frac{1}{2}||\mathbf{d} - \mathbf{Gm}||_2^2 $ requires the solution of a regularized least-squares inverse problem. Apart from some specific cases involving orthogonal operators, where an analytical solution can be identified (e.g., CT scan reconstruction - \cite{Corona}), the evaluation of such an operator is generally very expensive and requires the use of iterative solvers (e.g., CGLS, LSQR). However, as each iteration of the Primal-Dual algorithm represents only a step towards the saddle point of the corresponding Primal-Dual functional, an approximated solution of the problem obtained with an early stopping after a limited number of iterations generally suffices. Moreover, a warm start is generally employed, meaning that the solution of the proximal operator from its previous evaluation is used as starting guess of its current evaluation. This often provides a very large speed improvement over solving the problem from scratch each time \cite{Parikh}. 

Finally the proximal operator of the unit Simplex used within the segmentation process (equation \ref{eq:alg}c) also requires only point-wise evaluations, where each column of the matrix $\textbf{V}$ is treated independently and projected over the intersection of a hyperplane and a box. This is implemented as the evaluation of a projection over a box:
\begin{equation}
\label{eq:simplexprojection}
P_C = P_{Box_{[0, +\inf]}} (\mathbf{x} - \mu^* \mathbf{1})
\end{equation}
where $\mu$ is the solution of the following scalar equation
\begin{equation}
\label{eq:simplexbisect}
f(\mu) = \mathbf{1}^T P_{Box_{[0, +\inf]}} (\mathbf{x} - \mu \mathbf{1}) - 1
\end{equation}
which can be solved via bisection. The cost of this operation is dependant on the speed of convergence of the bisection algorithm and may require tens of evaluations of the function $f$. Note that, although this has a higher cost compared to the proximal operator of the $L_{2,1}$ norm, its implementation can be made very fast when implemented on modern, massively parallel hardware (i.e., GPUs). 

\subsection*{Extension to pre-stack data}
The algorithm presented in this work can be easily adapted to the case of pre-stack seismic data with the possibly added benefit of better separation between the different classes chosen as input to the segmentation step. As far as the inversion step is concerned, the model vector $\textbf{m}$ can simply be defined as a vector of size $N_m N_x N_z \times 1$ where $N_m$ is the number of model parameters we wish to invert for (generally 2 or 3). Similarly, we can define the data vector $\textbf{d}$ as a vector of size $N_\theta N_x N_z \times 1$ where $N_\theta$ is the number of angles. Finally, the modelling operator is redefined as $\textbf{G}=\textbf{WMD}$ to include a mixing matrix $\textbf{M}$ which contains the weighting coefficients of the different elastic reflectivities used to model the amplitude variation with offset (AVO) seismic response - e.g., Aki-Richards. Similar to the matrix $\textbf{V}$, we can also consider a matrix $\textbf{M}$ of size $N_x N_z \times N_m$ (where $\mathbf{m} = Vec(\mathbf{M})$ is its vectorized version obtained by concatenating the matrix's columns) and replace the TV regularization term in equation \ref{eq:joint} by the sum of $N_m$ TV norms acting on each column of $\textbf{M}$. The proximal operator of this term can therefore be computed in the same way as that of $\textbf{V}$ (equation \ref{eq:proxL21}). Alternatively, as described in \cite{Causse, Ravasi}, the different model parameters can be first decoupled from each other, followed by $N_m$ independent steps of inversion that are identical to the one employed in the post-stack scenario. 

Finally, for the segmentation step we can define a matrix $\mathbf{C}$ of size $N_c \times N_m$ whose rows contain the set of elastic parameters associated to each class and replace the $(m_i - c_j)$ in equation \ref{eq:joint} by the Euclidean norm of the difference between each row of $M$ and $||\mathbf{M}_i - \mathbf{C}_j||_2$. 

\section*{Conclusions}
In this work we have presented an end-to-end approach to assisted seismic interpretation which provides alongside a set of interpreted horizons also a property model (or a set of properties models for the elastic case) and a segmented model of the subsurface. Acoustic (or elastic) properties are not only inverted for the entire model from seismic data, but they are also used as input to define an easy to understand link between well log responses and the expected horizons to be extracted in the final step of our workflow. This in turn reconciles classical principles of seismic interpretation to the proposed optimization problem, making the method easier to explain to and be used by skilled interpreters. Our work remarks on the importance of enforcing blockiness in the inversion of seismic data, which is obtained via a combination of regularization (TV regularization), solver selection (Primal-Dual) and additional constraints (segmentation model): our estimates present sharper jumps at model discontinuities than their least-squares counterpart. This not only helps the subsequent segmentation step, but it has strong implications when these models are used as hard or soft constraints in downstream processes such as facies or property modelling. Finally, whilst an extension of our inversion algorithm to pre-stack seismic data has been discussed, we foresee its applicability to other linear (and nonlinear) inverse problems in geophysics, such as least-squares migration and waveform inversion.

\bibliographystyle{unsrt}  
\bibliography{references}

\end{document}